\documentclass[amsmath,amssymb,11pt]{article}
\usepackage{jheppub2}
\pdfoutput=1

\usepackage{amsmath,amssymb,amsthm,mathrsfs,bbm}
\usepackage{latexsym,amscd,amsbsy,amsfonts,dsfont}
\usepackage{graphicx}
\usepackage{xcolor}
\usepackage{caption}

\usepackage{chngpage} 

\usepackage[normalem]{ulem} 

\newcommand{\beq}{\begin{equation}}
\newcommand{\eeq}{\end{equation}}
\newcommand{\beqa}{\begin{eqnarray}}
\newcommand{\eeqa}{\end{eqnarray}}
\newcommand{\bea}{\begin{eqnarray}}
\newcommand{\eea}{\end{eqnarray}}

\newcommand{\ie}{{i.e.,\ }}
\newcommand{\eg}{{e.g.,\ }}

\newcommand{\lp}{\left(}
\newcommand{\rp}{\right)}

\newcommand{\ord}[1]{{\mathcal O}\lp #1\rp}

\newcommand{\T}{\mathbb{T}}

\newcommand{\N}{\mathbb{N}}

\newcommand*{\affmark}[1][*]{\textsuperscript{#1}} 

\numberwithin{equation}{section}  
\allowdisplaybreaks

	\usepackage{graphicx}
	\usepackage{tikz}
	\usetikzlibrary{decorations.markings}
	\tikzset{->-/.style={decoration={
				markings,
				mark=at position #1 with {\arrow{stealth}}},postaction={decorate}}}
	\usepackage{adjustbox}
	\usepackage{pgfplots}
	\pgfplotsset{compat=1.11}
	\usepgfplotslibrary{fillbetween}
	\usetikzlibrary{intersections}
	\usetikzlibrary{patterns}
	\usetikzlibrary {arrows.meta}
	\pgfdeclarelayer{bg}
	\pgfsetlayers{bg,main}
	
	\usetikzlibrary{calc}
	\tikzset{
		samples=100,
	}
	\pgfplotsset{compat=1.11}
	
	\pgfmathsetmacro\T{3.14}
	\pgfmathsetmacro\A{0.2}
	\pgfmathsetmacro\N{4}
	\pgfmathsetmacro\D{\N*\T}
	
\makeatletter
\pgfdeclareradialshading[tikz@ball]{ball}{\pgfqpoint{-10bp}{10bp}}{%
 color(0bp)=(tikz@ball!20!white);
 color(13bp)=(tikz@ball!25!white);
 color(18bp)=(tikz@ball!100!black);
 color(25bp)=(tikz@ball!70!black);
 color(50bp)=(black)}
\makeatother

\usetikzlibrary{fadings}
\tikzfading 
[
  name=fade out,
  inner color=transparent!0,
  outer color=transparent!100
]

\title{{{The correspondence between rotating black holes and fundamental strings}}}
\subheader{\begin{flushright}
\texttt{CPHT-RR037.072023}
\end{flushright}}

\author{Nejc \v{C}eplak,\affmark[1,2]}
\emailAdd{ceplakn@tcd.ie}
\author{Roberto Emparan,\affmark[3,4]}
\emailAdd{emparan@ub.edu}
\author{Andrea Puhm,\affmark[5,6]}
\emailAdd{a.puhm@uva.nl}
\author{Marija Tomašević\affmark[5,6]}
\emailAdd{m.tomasevic@uva.nl}

\affiliation{
\affmark[1]Universit\'e Paris Saclay, CNRS, CEA,\\
 Institut de Physique Théorique, 91191, Gif-sur-Yvette, France\\
 \affmark[2]School of Mathematics and Hamilton Mathematics Institute,
Trinity College, Dublin 2, Ireland\\
\affmark[3]Institució Catalana de Recerca i Estudis Avançats (ICREA)\\
 Passeig Lluis Companys, 23, 08010 Barcelona, Spain\\
\affmark[4]Departament de Física Quàntica i Astrofísica and
  Institut de Ciències del Cosmos,\\
 Universitat de Barcelona, 08028 Barcelona, Spain\\
\affmark[5]Centre de Physique Théorique (CPHT), Ecole Polytechnique, \\
Bâtiment 6, Route de Saclay, 91128 Palaiseau, Cedex, France\\
\affmark[6]Institute for Theoretical Physics, University of Amsterdam,\\ PO Box 94485, 1090 GL Amsterdam, The Netherlands}

\abstract{
The correspondence principle between strings and black holes is a general framework for matching black holes and massive states of fundamental strings at a point where their physical properties (such as mass, entropy and temperature) smoothly agree with each other. This correspondence becomes puzzling when attempting to include rotation: At large enough spins, there exist degenerate string states that seemingly cannot be matched to any black hole. Conversely, there exist black holes with arbitrarily large spins that cannot correspond to any single-string state. We discuss in detail the properties of both types of objects and find that a correspondence that resolves the puzzles is possible by adding dynamical features and non-stationary configurations to the picture. Our scheme incorporates all black hole and string phases as part of the correspondence, save for one outlier which remains enigmatic: the near-extremal Kerr black hole. Along the way, we elaborate on general aspects of the correspondence that have not been emphasized before. 
}

\begin{document}

\maketitle

\section{Introduction}
\label{sec:Intro}

One of the earliest suggestions that a relationship might exist between fundamental strings and black holes \cite{Green:1987sp} arose from the observation that, in both cases, their angular momentum is bounded above as 
\begin{equation}\label{JlM2}
    J\leq M^2\,.
\end{equation}
As we will presently see, this observation is misguided. Nevertheless, it provides a convenient starting point to understand how the correspondence between black holes and fundamental strings, as proposed in \cite{Susskind:1993ws,Horowitz:1996nw}, acquires new richness and intricacy when rotation is present.

The first problem with \eqref{JlM2} is that it conflates two very different bounds. This can be seen by restoring the units in which each of them is expressed.\footnote{This point is implicit in \cite{Bardeen:1999px}.} The Regge bound on the spin of string states is 
\begin{equation}\label{regge}
    J\leq \frac{M^2}{M_{s}^2}\,,
\end{equation}
where $M_s$ is the string mass scale, while the Kerr bound on the spin of a black hole is
\begin{equation}\label{kerr}
    J\leq \frac{M^2}{M_{P}^2}\,,
\end{equation}
where $M_P$ is the Planck mass (the numerical factors entering in the definitions of $M_s$ and $M_P$ are unimportant for our purposes, as will become clear). In general, $M_s$ and $M_P$ can be very different, so the two bounds can differ widely. These two mass scales are related (in four dimensions) through the string coupling constant $g$ as 
\begin{equation}
M_s = g \,M_P\,.
\end{equation}
This implies that when $g\ll 1$ and the strings are weakly coupled, the Kerr bound is saturated at much lower spins than the Regge bound,
\begin{equation}\label{kerrblessreggeb}
    J_\textrm{Kerr}=\frac{M^2}{M_{P}^2}= g^2\frac{M^2}{M_{s}^2}\ll J_\textrm{Regge}=\frac{M^2}{M_{s}^2}\,.
\end{equation}
This separation is relevant since, as we will review, the correspondence between massive string states with $M\gg M_s$ and black holes is indeed expected to occur at small values of the coupling. Namely, as $g$ becomes smaller, the curvature near the black hole horizon grows and it reaches the string scale when
\begin{equation}\label{gcorr}
    g^2\big|_\textrm{corr} \sim \frac{M_s}{M}\,.
\end{equation}
At this point,
\begin{equation}
  J_\textrm{Kerr}\Big|_\textrm{corr} \sim \frac{M}{M_{s}}\ll J_\textrm{Regge} \,.
\end{equation}
So, at first sight, \eqref{kerrblessreggeb} suggests that weakly coupled, highly massive string states with large enough spins in the range
\begin{align}
    g^2\left(\frac{M}{M_s}\right)^2<J<\left(\frac{M}{M_s}\right)^2
\end{align}
do not have any black hole counterpart to correspond to.

The second reason why a correspondence through \eqref{JlM2} is problematic is that the string states that saturate the Regge bound look nothing like black holes. Instead of balls of string, they resemble thin, long, rigidly rotating rods. Unlike the typical very massive string states with low angular momentum, these Regge-saturating states have very small degeneracy. Instead, extremal Kerr black holes are almost round and possess a large Bekenstein-Hawking entropy. Therefore, the states that saturate the two bounds appear to be completely unrelated to each other. Nevertheless, \eqref{kerrblessreggeb} says that, at weak coupling, the Kerr bound is saturated at angular momenta well below the Regge bound. Strings with spins in this range are still highly degenerate, so, in principle, a correspondence seems possible with black holes up to the latter's maximum angular momenta.\footnote{See \cite{Russo:1994ev} for an early study.}

This brings us to the third reason why \eqref{JlM2} is seemingly irrelevant to establish a correspondence. The upper bound on the spin \eqref{kerr} is valid for the four-dimensional Kerr black hole, but in five dimensions there exist rotating black rings with arbitrarily large spins for a given mass \cite{Emparan:2001wn}. In six or more dimensions, the generalizations of the Kerr solution found by Myers and Perry (MP) \cite{Myers:1986un} also possess such ultraspinning regimes. In contrast, the Regge bound on the spin of string states holds in every dimension. This bound must be reassessed for a possible correspondence where $g$ varies, but in any case, ultraspinning black holes and black rings do not look at all like rotating rods. We now face a problem opposite to the one we encountered earlier: There may not exist string states that are smoothly connected to these black holes. 

Thus we are confronted with two reciprocal puzzles: 
\begin{itemize}
    \item Imagine starting with a fast-spinning free string, then increasing $g$, making its gravitational self-interaction stronger. What does the string turn into?
    \item Conversely, suppose we begin with an ultraspinning black hole and decrease $g$ to a very low value. What does the black hole transform into?
\end{itemize} 

In this article, we will find answers to these questions within the broader correspondence between generic rotating neutral black holes and fundamental string states. As we will see, behind the puzzles posed above there lie hidden assumptions, most notably that the correspondence must exist as a one-to-one relation between stationary objects, stable or not, identified at both ends of the range of the varying coupling $g$. This relation, however, can never be entirely correct, since once the coupling $g$ takes on a finite, non-zero value, interactions render all of the states involved unstable and time-evolving. 
Usually, these effects are tacitly assumed to be negligible, since the decay rates of both massive strings and black holes at the correspondence are slow. However, they are necessarily non-zero, and a notion of adiabaticity relating states at different values of $g$ does not strictly exist.
For static black holes, one can formulate a suitable approximation of `Goldilocks adiabaticity', but when rotation is present this will only be satisfied by some states and not by others. As a result, the picture that we find involves considerably richer physics than for non-rotating objects, and in particular, dynamical factors are crucial to establishing the correspondence.

We will elaborate in detail on these considerations to arrive at a comprehensive account of the correspondence for all neutral black holes (and dipole black rings) with a single angular momentum in $D\geq 4$---with only one significant exception: the near-extremal four-dimensional Kerr black hole.\footnote{The picture extends to more angular momenta without any significant qualitative changes, but also with near-extremal multi-spin outliers.} The self-gravitation of the string, which in general plays a role in completing the details of the correspondence, may be crucial for understanding the stringy nature of near-extremal Kerr throats. 

\begin{figure}
        \centering
         \includegraphics[width=.7\textwidth]{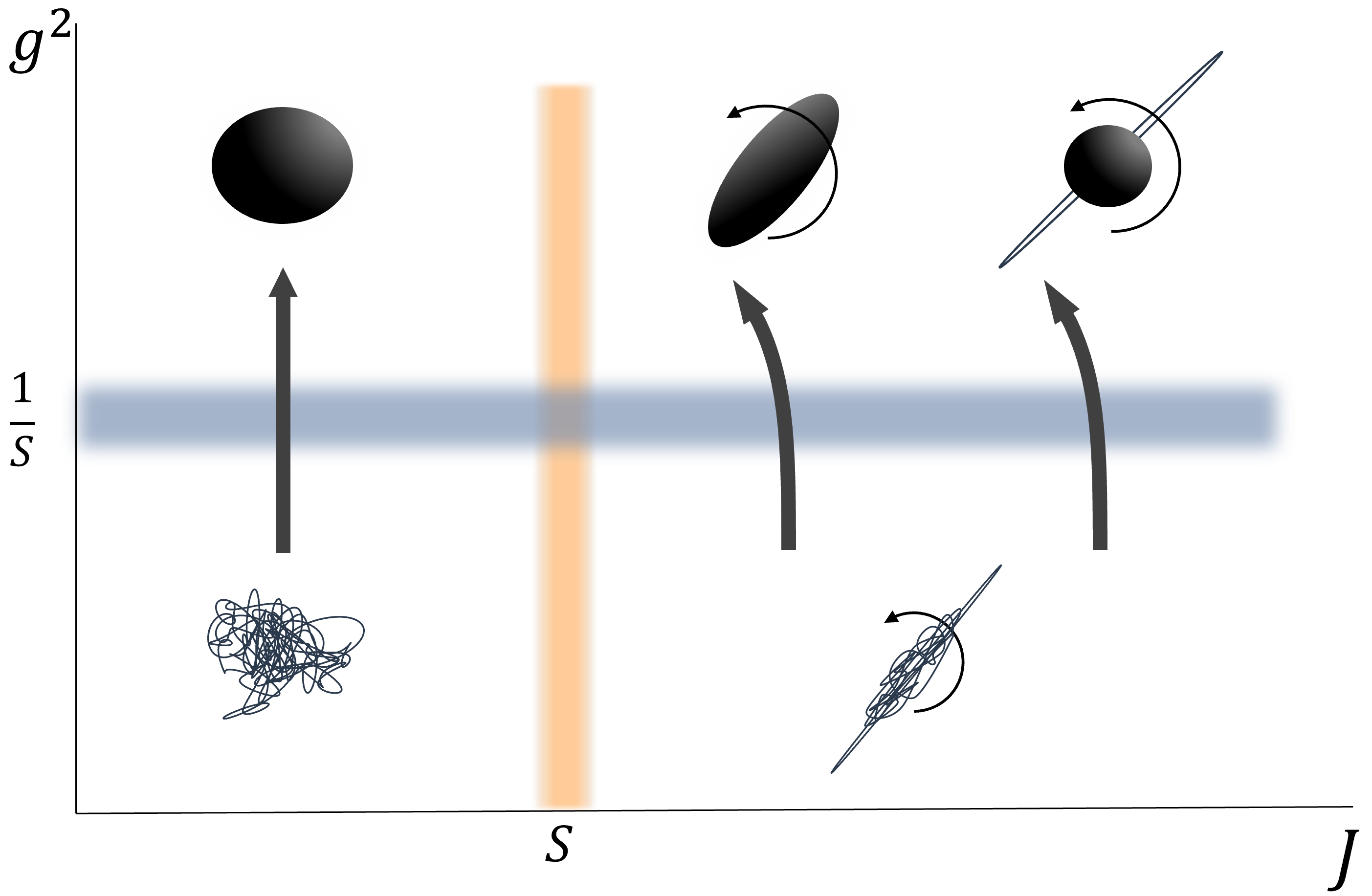}
        \caption{\small Correspondence from string states to black holes for different values of $J$. The entropy $S$ of the states is fixed, and we follow them as $g$ is increased, starting as free string states at $g=0$ and adiabatically collapsing them into black holes when $g^2=1/S$ (horizontal band). The vertical band marks the separation between the regime of stable Kerr-like black holes with $J<S$ and the ultraspinning regime with $J>S$. Order-one numerical factors are ignored throughout. In the left side of the diagram, spinning string balls with $J<S$ evolve into Kerr-like black holes much as they do without rotation. To the right, with $J>S$, typical string states are rotating `wiggly rods', which transition into either rotating black bars or black hole-string hybrids (the former more plausibly in $D\geq 5$, the latter in $D=4$). None of these are stationary solutions and will spin down through radiation emission until they settle into a Kerr-like black hole. However, the spin loss can be slow enough to allow an approximately adiabatic transition at the correspondence. Self-gravity introduces a dependence on dimension in the correspondence which we have not incorporated here.}
        \label{fig:StrBH}
\end{figure}

\begin{figure}
        \centering
         \includegraphics[width=\textwidth]{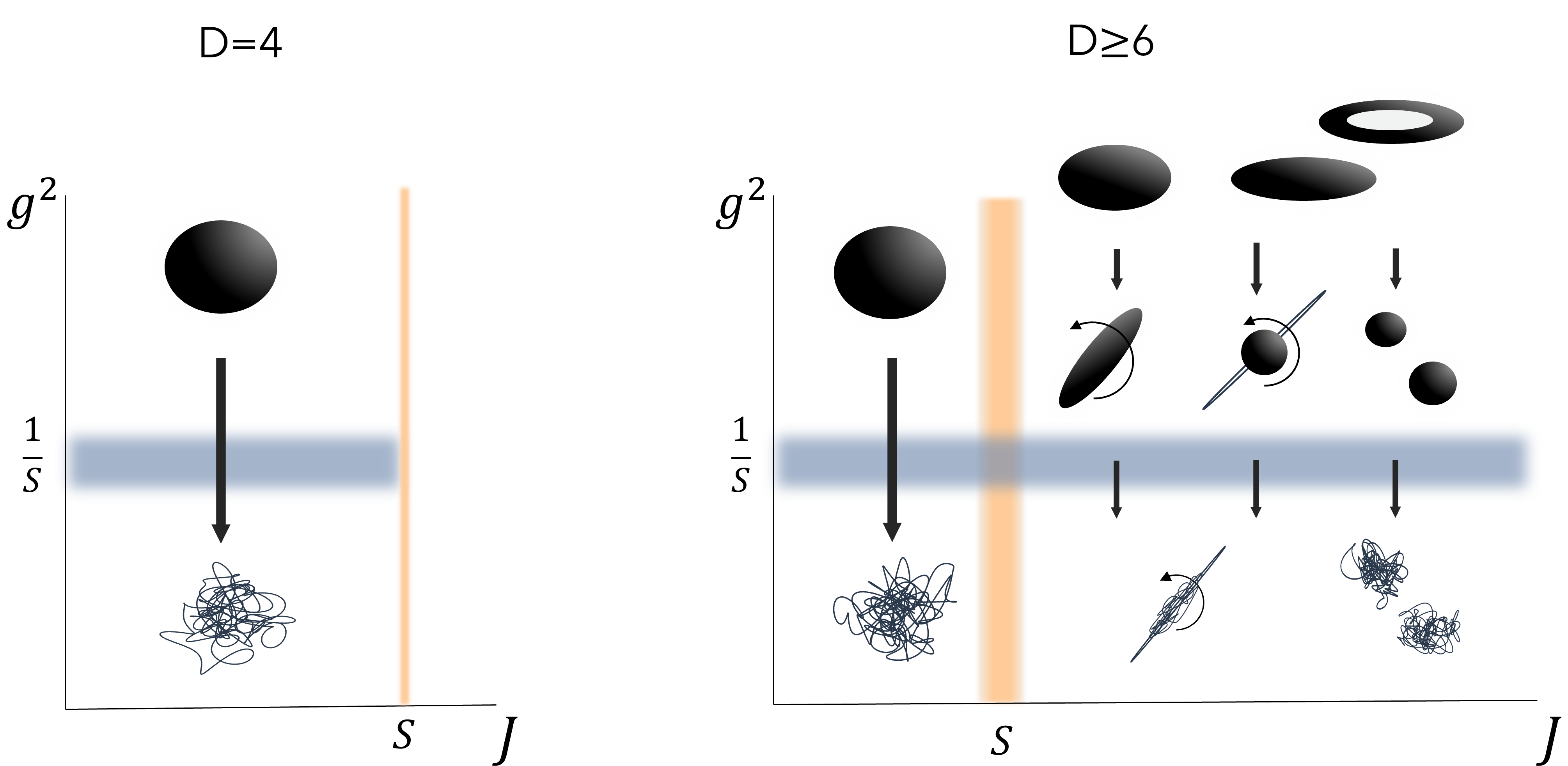}
        \caption{\small Correspondence from black holes to string states. We start at a large value of the coupling and reduce it until the horizon curvature reaches the string scale. The black hole then transitions into a string state parametrically smoothly connected to it. Left: in $D=4$, Kerr black holes evolve into slowly spinning string balls. When the black hole is near the extremal limit (the precise value is $S=2\pi J$) the transition into a string ball is not adiabatic, and the nature of the correspondence is unclear. Right: in $D\geq 6$, Kerr-like black holes with $J<S$ are classically stable and evolve as in $D=4$. Ultraspinning black holes and black rings with $J>S$ have fast dynamic instabilities that affect them before they reach the correspondence. For moderately large spins, bar instabilities dominate and the black bars transition into stringy wiggly rods. For larger spins, fragmentation instabilities break up the black hole into smaller Kerr-like black holes that then evolve into several string balls. Black hole-string hybrids may also appear as intermediate states. The picture in $D=5$ shares many similarities with $D\geq 6$, but hybrids play a more prominent role (figure~\ref{fig:BHStr5}). Another relevant correspondence involves plasmid strings and dipole black rings (section~\ref{sec:dipoleplasmids}). }
        \label{fig:BHStr46}
\end{figure}

The main elements in our picture are summarized in figures~\ref{fig:StrBH} and \ref{fig:BHStr46}, while the detailed discussion is given in section~\ref{sec:Correspondence}.\footnote{To generate the illustrations of string balls we have followed \cite{Karliner:1988hd}.}

We begin in the next section by revisiting the correspondence principle as formulated in \cite{Susskind:1993ws,Horowitz:1996nw,Susskind:2021nqs}, emphasizing aspects that have not been explicitly made before but which will be relevant for us. Afterward, we introduce the main characters in the correspondence. First, in section~\ref{sec:bhs} we briefly review the essential features of rotating black holes in different dimensions (with rotation on a single plane) and discuss their instabilities and decays. Then, in section~\ref{sec:stringstates} we turn to describe the construction of spinning string states, especially those with large degeneracies, which have been scarcely studied earlier. An adequate understanding of the properties of these objects will then allow us, in section~\ref{sec:Correspondence}, to establish the main elements of the correspondence for neutral black holes and strings in different dimensions. Section~\ref{sec:dipoleplasmids} explains how the correspondence picture also applies for rotating massive states with a fundamental-string dipole.  The extension to multiple angular momenta is qualitatively simple, and we describe it in section~\ref{sec:MultipleJ}. In section~\ref{sec:evapcorr} we discuss the correspondence as a physical process at a late stage in black hole evaporation, and we comment on the role of the Page time in the corresponding string state. Section~\ref{sec:redux} reexamines the nature of the ensembles of states that are related by the correspondence, making a comparison to how it works for BPS systems. Finally, in section~\ref{sec:Summary} we summarize how the puzzles we posed above are resolved, and highlight open issues.

\section{Correspondence for non-rotating objects}
\label{sec:ReviewStatic}

\subsection{Basic picture}

Let us begin with a static, neutral black hole in $D$ dimensions. Its mass $M$ defines a length scale that sets the horizon radius to
\begin{align}\label{rhM}
    r_H = \left(G M\right)^{\frac{1}{D-3}}= \left(\frac{M}{M_P}\right)^{\frac{1}{D-3}}\,\ell_P\,.
\end{align}
We have introduced here the Planck mass and Planck length which (since we always set $\hbar =1$) replace Newton's constant $G$ as
\begin{align}
    G = M_P^{-D+2} = \ell_P^{D-2}\,.
\end{align}
Throughout the article, we use the equal sign even if we do not retain any precise numerical factors unless necessary (this is only in sec.~\ref{sec:stringstates} and partly in sec.~\ref{subsec:BHSt5D}). Then, we can easily keep track of scales by simple dimensional arguments.

In string theory, the Planck scale is not fundamental but a derived quantity. The tension of the strings, denoted for historical reasons as $1/2\pi \alpha'$, sets the fundamental mass and length scales as
\begin{align}
    \frac1{2\pi\alpha'}=M_s^2=\ell_s^{-2}\,.
\end{align}
The probability amplitude that two segments of closed string within a distance $\ell_s$ reconnect is the string coupling $g$. This sets the strength of string interactions, including gravitational ones, which implies that
\begin{align}\label{Gells}
    G=\ell_P^{D-2}=g^2 \ell_s^{D-2}\,.
\end{align}
Using now string units, the radius of the horizon is
\begin{align}\label{rHls}
    r_H=\lp g^2 \frac{M}{M_s}\rp^{\frac1{D-3}}\ell_s\,,
\end{align}
so the 
curvature near the black hole, measured e.g., by the Kretschmann scalar, is
\begin{align}\label{Kbh}
    \mathcal{K}\equiv R_{abcd}R^{abcd}= \frac1{r_H^4}=\lp\frac{M_s}{g^2 M}\rp^{\frac4{D-3}}\frac1{\ell_s^4}\,.
\end{align}
We can see that the curvature increases as we reduce $g$.

The string coupling $g$ is actually not a constant but can vary in space and time, being given by the expectation value of the dilaton field $\phi$,
\begin{align}
    g=e^\phi\,.
\end{align}
We will be interested in slowly varying $g$ between small and large values, which can be done by altering the value of dilaton field. 
It is possible to imagine doing this in a physical way by letting a dilaton wave of a very long wavelength pass by the spatial region of interest. 
At low energies, string theory is well described in terms of an effective action of the schematic form (neglecting indices) \cite{Callan:1988hs}
\begin{align}\label{Istframe}
    I_{\rm{eff}}=\frac1{\ell_s^{D-2}}\int \sqrt{-\det g_{ab}}\,e^{-2\phi}\lp R + \ell_s^2 R^2 +\dots\rp\,,
\end{align}
where we ignore the kinetic term for the dilaton (and other fields) since we will assume that its gradients have a negligible effect on the metric. Here $R$ is the curvature of the string metric $g_{ab}$ (the one to which strings minimally couple), and the dots denote higher-order stringy corrections. We see that varying $\phi$ has the effect of varying the gravitational coupling, so that $G\sim e^{2\phi}\ell_s^{D-2}$, as we already saw in \eqref{Gells}.

\begin{figure}
        \centering
         \includegraphics[width=.7\textwidth]{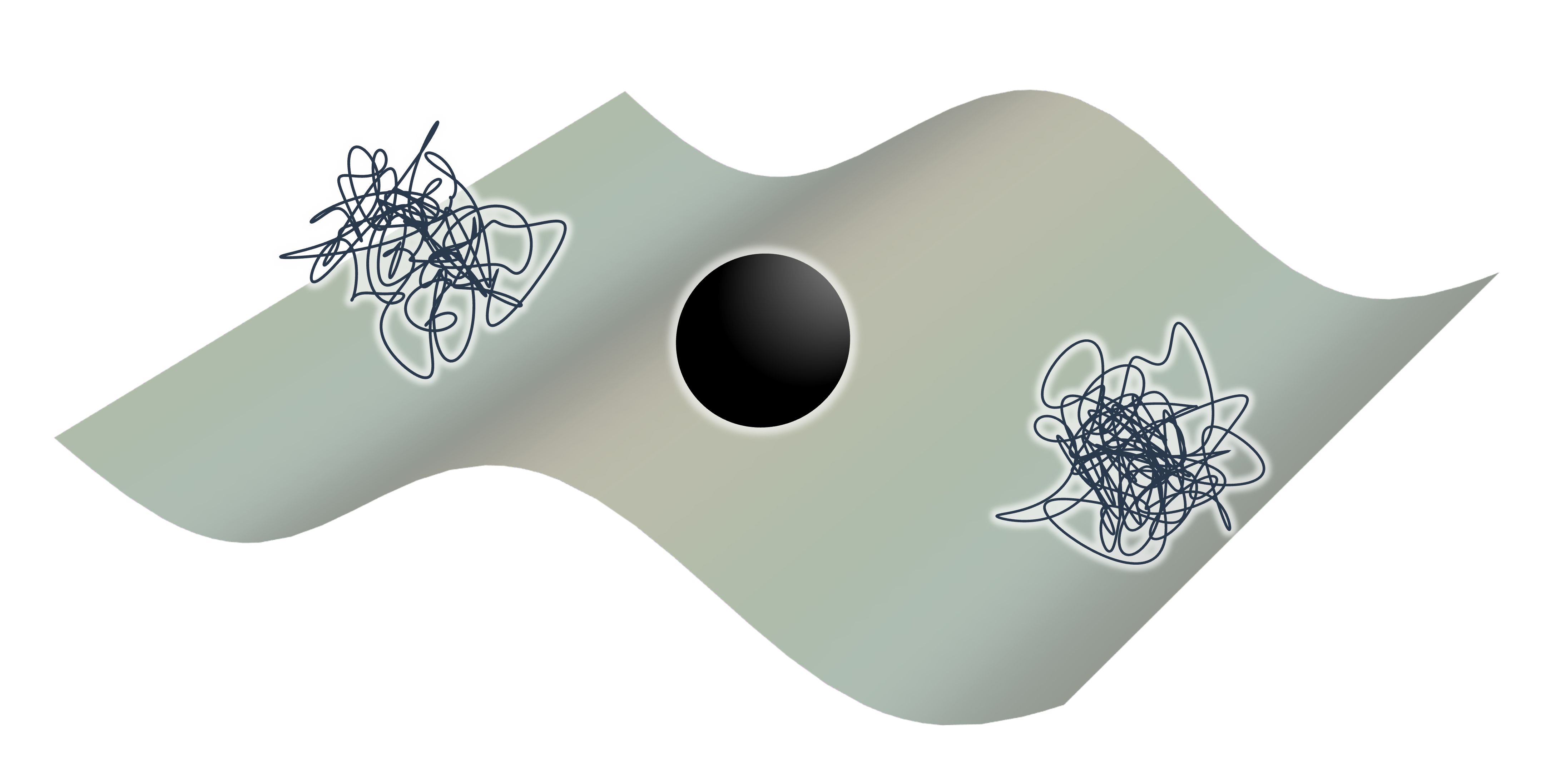}
        \caption{\small As the dilaton wave passes, it locally changes the value of the string coupling and the system alternates between the form of a black hole and a massive string. The coupling oscillates around the value $g^2=1/S$ and, for an adiabatic change, the wavelength must be much longer than the string scale $\ell_s$.}
        \label{fig:wave}
\end{figure}

For a black hole, the natural metric (the one where its horizon satisfies the area theorem) is not $g_{ab}$ but the Einstein-frame metric
\begin{align}
    g_{ab}^E =e^{-4\phi/(D-2)}g_{ab}\,,
\end{align}
in terms of which\footnote{We assume for simplicity that the dilaton vanishes asymptotically.}
\begin{align}\label{Ieiframe}
    I_{\rm{eff}}=\frac1{\ell_s^{D-2}}\int \sqrt{-\det g_{ab}^E}\lp R_E + \ell_s^2 e^{-\frac{4\phi}{D-2}} R_E^2 +\dots\rp\,.
\end{align}
In this frame, the dilaton does not directly couple to the curvature to leading order. Therefore, as the dilaton wave passes by a neutral black hole, none of the properties of the black hole, such as its mass and area from the leading order gravitational action (hence in Planck units), will change. However, decreasing the value of $\phi$, that is, decreasing  $g$, enhances the stringy corrections to the geometry. Then, when $g$ becomes small enough, we expect the black hole to become a string-like object   (see figure~\ref{fig:wave}).

Indeed, when 
\begin{align}\label{gcorrM}
    g^2 = \frac{M_s}{M}
\end{align}
the horizon radius \eqref{rHls} becomes string size, $r_H=\ell_s$, and the correction terms in \eqref{Istframe} become non-negligible.\footnote{It is easy to verify that the Einstein-frame corrections in \eqref{Ieiframe} also become important at this moment, see appendix~\ref{app:einscorr}.} Since the Bekenstein-Hawking entropy of the black hole is
\begin{align}\label{SBHM}
    S=\frac{r_H^{D-2}}{G}=\lp \frac{M}{M_P}\rp^\frac{D-2}{D-3}=\frac1{g^2}\lp g^2\frac{M}{M_s}\rp^\frac{D-2}{D-3}\,,
\end{align}
then we can characterize \eqref{gcorrM} as the point where\footnote{A similar point was first made early on in \cite{Bowick:1985af}.}
\begin{align}\label{gcorrS}
    g^2 = \frac1{S}\,.
\end{align}
We will refer to this value of the string coupling---at which we expect the transition between a static black hole and a massive string state---as the \emph{correspondence point} \cite{Susskind:1993ws, Horowitz:1996nw}.

We see that the dimensionless parameter that controls whether we are in the stringy or in the black hole regime is not really $g^2$ but $g^2 S$. This is apparent if we write 
\begin{align}\label{KbhS}
    \mathcal{K}= \lp g^2 S\rp ^{-\frac4{D-2}}\frac1{\ell_s^4}\,.
\end{align}
When $g^2 S\gg 1$ the horizon curvature is much lower than the string scale, while for $g^2 S\lesssim 1$ we are in a fully stringy regime. In this respect, $g^2 S$ plays the same role as the 't~Hooft coupling $\lambda=g^2_{YM} N$ in AdS/CFT, or $g N$ in D-brane systems, with $N$ the number of D-branes\footnote{The different powers of $g$ reflect that fundamental string masses are $\propto g^0$ while D-brane masses are $\propto g^{-1}$.}. The classical black hole regime need not require $g\to\infty$, but rather $g^2 S\to \infty$, so semiclassical approximations to black holes remain valid when $g$ is small if $S$ is sufficiently large. This will be important in the physical setup we have described, where $S$ is fixed but always large, and we vary $g$ around the value \eqref{gcorrS} never needing to take $g\gg 1$.

Let us now consider highly excited states of a string. As we will explain in more detail later, their properties can be obtained to a very good approximation by considering them as very long classical strings with a random-walk profile of steps (or string bits) of length $\ell_s$ and mass $M_s$. The total mass of such a string is given by its number of string bits and is thus proportional to its length $L$,
\begin{align}
    M=\frac{L}{\ell_s}M_s\,.
\end{align}
For a given mass or total length, there are a large number of possible random walks. This means that the configurations for fixed mass have a large degeneracy since at each step the random walk can take any direction, which implies that the entropy of the string state is also proportional to its number of string bits,
\begin{align}
    S=\frac{L}{\ell_s}\,,
\end{align}
and therefore
\begin{align}\label{Sst}
    S=\frac{M}{M_s}\,.
\end{align}
This is in agreement with a rigorous calculation for quantized free strings, which yields a Hagedorn density of states $\rho(M)=e^{S(M)}\sim e^{M/M_s}$.

Now imagine sending a dilaton wave onto the string, with a wavelength long enough to neglect absorption or scattering. That is, we consider an adiabatic change of $g$, such that the entropy of the string state remains constant while the strength of the gravitational self-interaction slowly grows. The negative potential energy thus induced will reduce the total energy of the string state, or in other words, the mass of the string will be renormalized by the interactions. If the coupling $g$ is weak, we expect that this is a small effect,
\begin{align}\label{MSst}
    M= S M_s - \ord{g^2}\,.
\end{align}

\begin{figure}
        \centering
         \includegraphics[width=.5\textwidth]{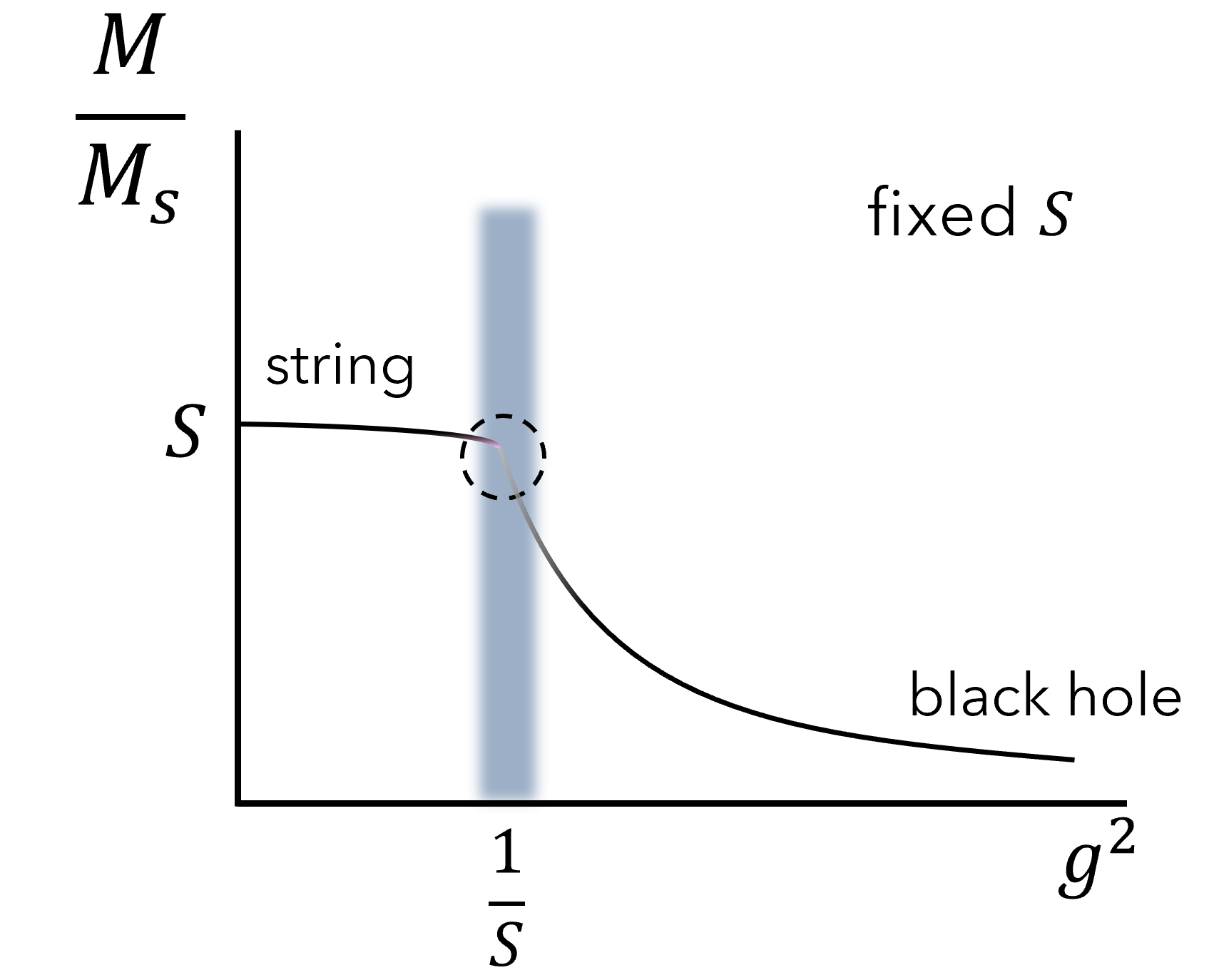}
        \caption{\small Correspondence between massive strings and static black holes along states of fixed entropy for varying string coupling $g$. The adiabats \cite{Susskind:2021nqs} for string states, $M/M_s=S-\ord{g^2}$, and for black holes, $M/M_s=g^{-\frac1{D-2}}S^{\frac{D-3}{D-2}}$, meet at a parametrically-defined band around the correspondence point $g^2=1/S$.}
        \label{fig:adiabat}
\end{figure}

If instead we send the wave onto a black hole and slowly reduce the coupling $g$, its entropy will remain fixed but, according to \eqref{SBHM}, its mass in string units will grow as
\begin{align}
    M =\frac{M_s}{g^2}\lp g^2 S\rp^\frac{D-3}{D-2}\,.
\end{align}
We can now see that, when the coupling reaches the value \eqref{gcorrS} where the black hole becomes a stringy object, its mass becomes $M\sim M_s S$, which parametrically agrees with that of a string state of the same entropy, \eqref{MSst} (see figure~\ref{fig:adiabat}).

That is, by changing the coupling $g$ adiabatically, we can smoothly follow a highly degenerate state with $S\gg 1$ and watch it oscillate between the form of a black hole and of a string state.\footnote{Ref.~\cite{Chen:2021dsw} finds that, once self-gravitation is included, for type II strings (but not for heterotic strings) there cannot be a connection as classical solutions of string theory.} Since the coupling at the correspondence point \eqref{gcorrS} is small, we can reliably use the leading order result in \eqref{MSst} as the statistical (coarse-grained) description of the string state, and thus provide a microscopic origin of the black hole entropy at the correspondence \cite{Susskind:1993ws,Susskind:2021nqs}.

\subsection{Sizes}\label{subsec:sizes}

As we will see in the next subsection, other properties such as the decay rates continuously match between the two descriptions of the states across the correspondence. However, there is an apparent discrepancy in their sizes, since a random-walk string spreads over a distance $\sqrt{L\ell_s}=\sqrt{S}\,\ell_s$, which is much larger than the horizon size of a black hole at the correspondence point, which is $\ell_s$. 

This might question the existence of a smooth transition from a string to a black hole (but not the other way around). It is clear that if the gravitational interaction grows large enough, the string will eventually collapse into a black hole. However, if this occurs only when the coupling is much larger than \eqref{gcorrS}, the entropy will discontinuously jump up in the collapse and the transition will not be adiabatic. Addressing the disparity requires properly accounting for the self-gravitation of the string state, which was done in \cite{Horowitz:1997jc} and \cite{Damour:1999aw}, and has been revisited recently in \cite{Chen:2021dsw,Brustein:2021cza,Matsuo:2022kvx,Urbach:2022xzw,Balthazar:2022hno, Urbach:2023npi}.  

The main effect of self-gravity is to counteract the tendency of the string ball to expand due to radial pressure and the centrifugal potential. These  decay like $\ell_s^2/r^2$ in all $D$, while, on the other hand, the gravitational self-attraction falls off as $ -g^2 M/r^{D-3}$. In $D=4$ this self-interaction is sufficiently long-ranged to gradually shrink the string ball as $g$ grows, until it reaches string-scale size at the correspondence point. In $D=5$ the effect is more abrupt, with the string ball quickly collapsing as \eqref{gcorrS} is approached, but still in a continuous way. In $D=6$, gravity is too short-ranged, with the effect that the random-walk string ball will not form a black hole until a larger value of $g$, at which point it collapses non-adiabatically  \cite{Horowitz:1997jc}. However, the authors of \cite{Damour:1999aw} claim that before the correspondence, the typical (i.e., most entropic) string states in $D=6$ are not random walks but more compact string balls, which transit adiabatically to black holes. In our opinion, there is still room for further clarification, but we take \cite{Damour:1999aw} as pointing to the possibility that, in $D\geq 6$, the large gravitational force at the short distances where the random walk is most dense, can have important effects.

In this article, we will bear in mind these issues for the evolution of strings to black holes, but without adding any further new detail. Self-gravitation of rotating string states will be considered elsewhere \cite{santoszigdon,usfollowup}.

\subsection{Decay rates and Goldilocks adiabaticity}\label{subsec:goldilocks}

Once the coupling $g$ takes on a finite non-zero value, none of the states considered above---strings or black holes---can be regarded as stationary anymore. Whenever $g\neq 0$, the self-interaction of a massive string state (i.e., the recombination of segments of it) leads to its decay by fragmentation or emission of light strings (radiation). Average string states of large mass are expected to decay dominantly through the latter. Fragmentation into separate massive string balls is possible, but so is the fusion of the balls too, and, as we saw, self-gravitation is likely to make the string ball more tightly bound and more compact \cite{Horowitz:1997jc}. 

Massless radiation emission happens with a thermal spectrum at the typical string-scale Hagedorn temperature \cite{Amati:1999fv}. It is also proportional to $g^2$ (for zero coupling the string is absolutely stable), so the decay rate is \cite{Iengo:2006gm,Damour:1999aw}
\begin{align}\label{Gammast}
    \Gamma\sim g^2 M = g^2 S\,\ell_s^{-1}\,.
\end{align}

On the other hand, turning on quantum interactions makes a black hole decay through the emission of Hawking radiation. The decay rate is governed by the Hawking temperature, which only vanishes when the black hole is infinitely larger than the Planck length. For us, with
\begin{align}\label{GammaBH}
    \Gamma\sim T_{BH}=\frac1{r_H}=\lp g^2 S\rp^{-\frac1{D-2}}\ell_s^{-1}\,,
\end{align}
this requires the 't~Hooft coupling $g^2 S\to\infty$. 

\begin{figure}
        \centering
         \includegraphics[width=.55\textwidth]{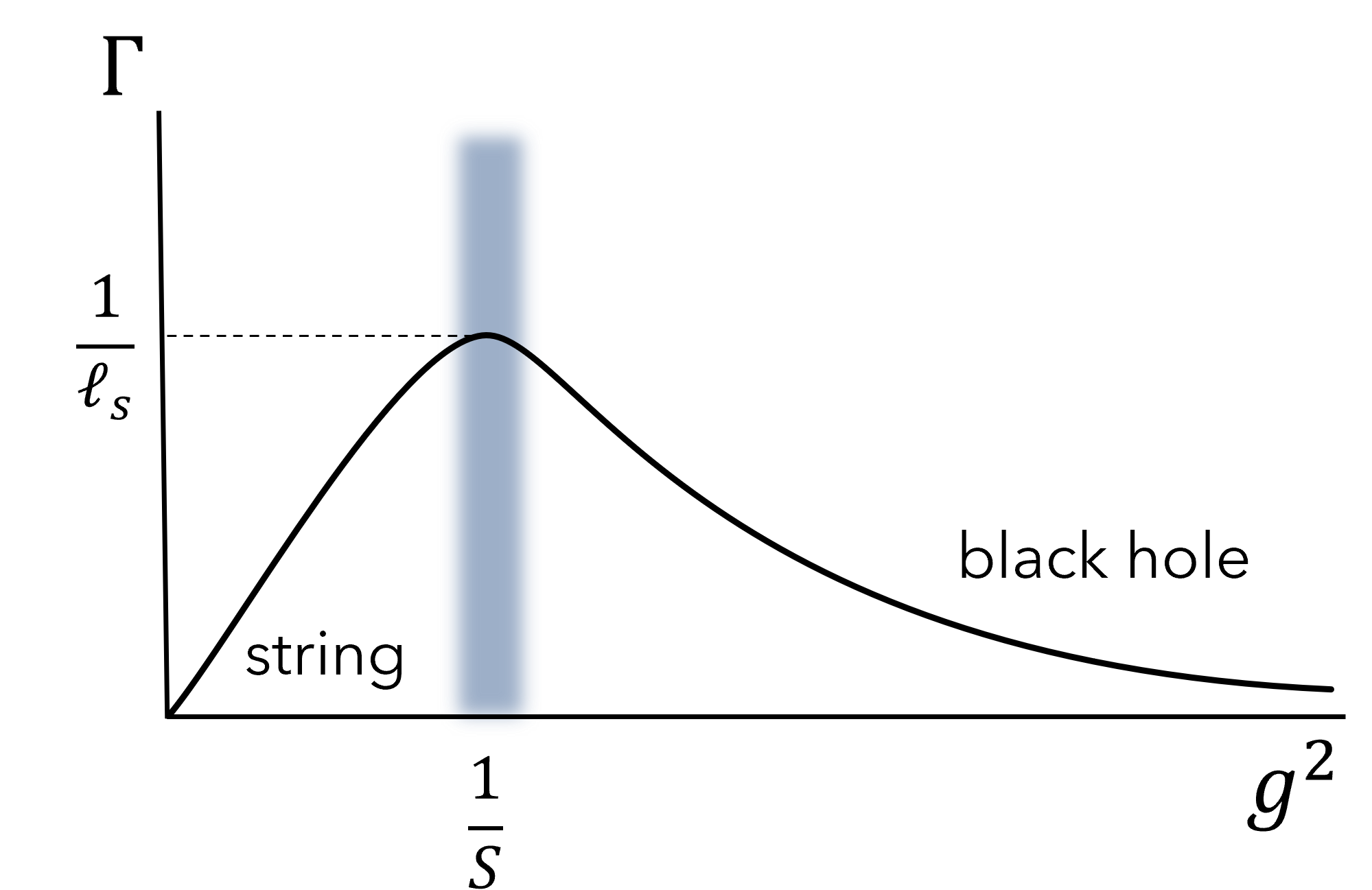}
        \caption{\small Decay rates $\Gamma$ of typical massive string states and static black holes, with fixed entropy and varying string coupling $g$. At the correspondence where $g^2=1/S$, they give a maximum string-scale decay rate.}
        \label{fig:decay}
\end{figure}

Therefore, the states we attempt to follow by varying $g$ are stationary only at the two ends $g^2 S\to 0$ and $g^2 S\to\infty$. If we work at large but finite $S$, this means $g\to 0$ and $g\to\infty$ (see figure~\ref{fig:decay}). The correspondence being at a finite and non-zero value of the  coupling prevents objects on either side of it from radiating with a very large decay rate. The rate reaches a maximum at the correspondence point \eqref{gcorrS}, where both \eqref{GammaBH} and \eqref{Gammast} give a string-scale thermal decay rate 
\begin{align}
    \Gamma_\mathrm{corr}=\ell_s^{-1}\,.
\end{align}

In light of this, let us now reexamine the assumption of adiabaticity along the correspondence transition. The usual condition of adiabaticity is that the rate of change
\begin{align}
    \Delta t_g^{-1} = \frac{\dot g}{g}=\dot\phi
\end{align}
should not be so fast as to excite the state under consideration. In terms of the dilaton wave, this requires that its wavelength be long enough that it is almost not absorbed (or radiated through stimulated emission) by the system. Since near the correspondence the size of the system is $\sim \ell_s$, adiabaticity requires that
\begin{align}\label{Gammacorr}
    \Delta t_g^{-1} < \frac1{\ell_s}\,.
\end{align}

The strict adiabatic limit is $\ell_s \Delta t_g^{-1}\to 0$. In our system, this limit can never be attained, since if the rate of change of $g$ is too slow, adiabaticity will be broken in a different way.
At any finite $g$ the system---either in the black hole or in the string phase---is emitting quanta and thus losing entropy. We must demand that the rate of change be fast enough to not radiate too many quanta across the transition. At a temperature $T$, a quantum is emitted every thermal time $1/T$, so the loss of entropy in an interval $\Delta t$ is
\begin{align}
    -\delta S = T \Delta t\,,
\end{align}
and we want to keep this loss small, $-\delta S < S$, that is,
\begin{align}
    \Delta t_g^{-1} > \frac{T}{S}\,.
\end{align}
As we decrease the string coupling, the Hawking temperature \eqref{GammaBH} increases until it reaches the Hagedorn temperature, $T=\ell_s^{-1}$, at the correspondence point. This maximal value gives a conservative estimate for the lower bound of the Goldilocks range for the rate of $g$---not too slow, not too fast:
\begin{align}\label{goldilocks}
    \frac1{S\ell_s}<\Delta t_g^{-1} < \frac1{\ell_s}\,.
\end{align}
Since we assume $S\gg 1$, there is ample range to satisfy this condition.\footnote{Away from the correspondence point, the Goldilocks range for a black hole is $\frac1{S r_H}<\Delta t_g^{-1} < \frac1{r_H}$ (with $r_H\geq \ell_s$), and for a string state it is $\frac{g^2}{\ell_s}<\Delta t_g^{-1} < \frac1{\ell_s}$ (with $g^2\leq 1/S$). 
These ranges shift with $g$, so away from the correspondence $\dot g$ must be adequately adjusted. But the Goldilocks conditions can always be comfortably met.} But we stress that, when $S$ is finite, it is impossible to strictly achieve the adiabatic limit of $\ell_s \Delta t_g^{-1}\to 0$.

\section{Rotating black holes}
\label{sec:bhs}

We now turn to examine the states of black holes and strings with non-zero net angular momentum. For the most part, we will consider that the rotation happens on a single plane out of the $\lfloor\frac{D-1}2\rfloor$ possible ones. This is not merely for simplicity but also because the most extreme consequences of rotation arise in this instance. Once we understand this case, we will be able to infer the main features of configurations with more general spins.

\subsection{Black hole phases}

We begin with a lightning review of the most salient properties of neutral black holes in all $D\geq 4$, with rotation on a single plane and with a connected horizon. It is convenient for our purposes to express results using the adiabatic invariants $S$ and $J$. This turns out to be illuminating. 

\subsubsection{Bird's eye view}

Rotating neutral black holes are conveniently characterized using the two classical length scales they possess \cite{Emparan:2009at}: the `mass length' and the `spin length',\footnote{In contrast to \eqref{rhM}, when the black hole rotates, $\ell_M$ does not correspond to the horizon radius.}
\begin{align}
    \ell_M=(GM)^\frac1{D-3}\,,\qquad \ell_J=\frac{J}{M}\,.
\end{align}
Then,
\begin{itemize}
    \item when $\ell_M > \ell_J$, there is a unique black hole, the Myers-Perry (MP) solution, of approximately round shape, which is dynamically stable. We will refer to this as the `Kerr regime';
    \item when $\ell_M < \ell_J$, there is a variety of black holes with different shapes and topologies, with horizons that are longer along the plane of rotation than in transverse directions, and which are dynamically unstable. This is the `ultraspinning regime'.
\end{itemize}
This simple characterization will be almost everything we need, but we must acknowledge that we are glossing over a great many details. First, in the demarcation of these two regimes, there is a $D$-dependent numerical factor that we are ignoring, and which in general is not very precisely known nor defined (e.g., the uniqueness bounds and stability bounds are close but not exactly equal). When $D$ is not very large, this number is always $\ord{1}$.

Second, there are relevant differences between $D=4$, $D=5$ and $D\geq 6$. In four dimensions, stationary black holes only exist in the Kerr regime, and the bound on the spin $\ell_J \leq \ell_M$ is saturated by extremal black holes, which remain round and smooth but have zero temperature. In five dimensions, there is an extremal limit for the MP black hole, but it is a singular solution
with zero horizon area. 
In five dimensions there exists an ultraspinning regime with black rings that are dynamically unstable \cite{Emparan:2001wn,Elvang:2006dd,Santos:2015iua,Figueras:2015hkb}. 

The landscape in $D\geq 6$ is qualitatively the same across all $D$: there is no extremal limit for the MP solutions, but rather the bound
\begin{align}\label{kerrb1}
    \ell_J\leq \ell_M \qquad \textrm{(`Kerr bound')}
\end{align}
is a dynamical stability limit \cite{Emparan:2003sy,Dias:2009iu,Shibata:2010wz,Dias:2014eua,Bantilan:2019bvf}. Nevertheless, we will still refer to it as the Kerr bound. 

In $D\geq 6$, for spins above the dynamical Kerr bound there exist ultraspinning MP black holes, black rings, and also bumpy black holes that connect the former two---all of them dynamically unstable. Black rings typically have the highest entropy for a given mass and spin \cite{Emparan:2007wm,Dias:2014cia,Emparan:2014pra}\footnote{This is intuitively clear: the energy of the black hole consists of heat (entropy) and rotational energy. Since a ring has a larger moment of inertia than a disk, it uses up less rotational energy to carry a given spin, thus leaving more energy to heat.}. 

In $D\geq 5$, several solutions coexist around the regime of $\ell_M \sim \ell_J$---fat and thin black rings, MP black holes, and bumpy black holes---and their relative dominance in entropy and stability is a complex affair that we will not attempt to resolve. The main qualitative features will suffice for us.

\subsubsection{Fixed-entropy diagrams}\label{subsubsec:fixedS}

These phases of black holes are often presented in diagrams of $S$ as a function of $J$ for fixed $M$ \cite{Emparan:2008eg}. However, for us it is more convenient to fix the entropy and then represent $M$ as a function of $J$ (both in Planck units). These diagrams reveal simple general patterns.

\begin{figure}
        \centering
         \includegraphics[width=.45\textwidth]{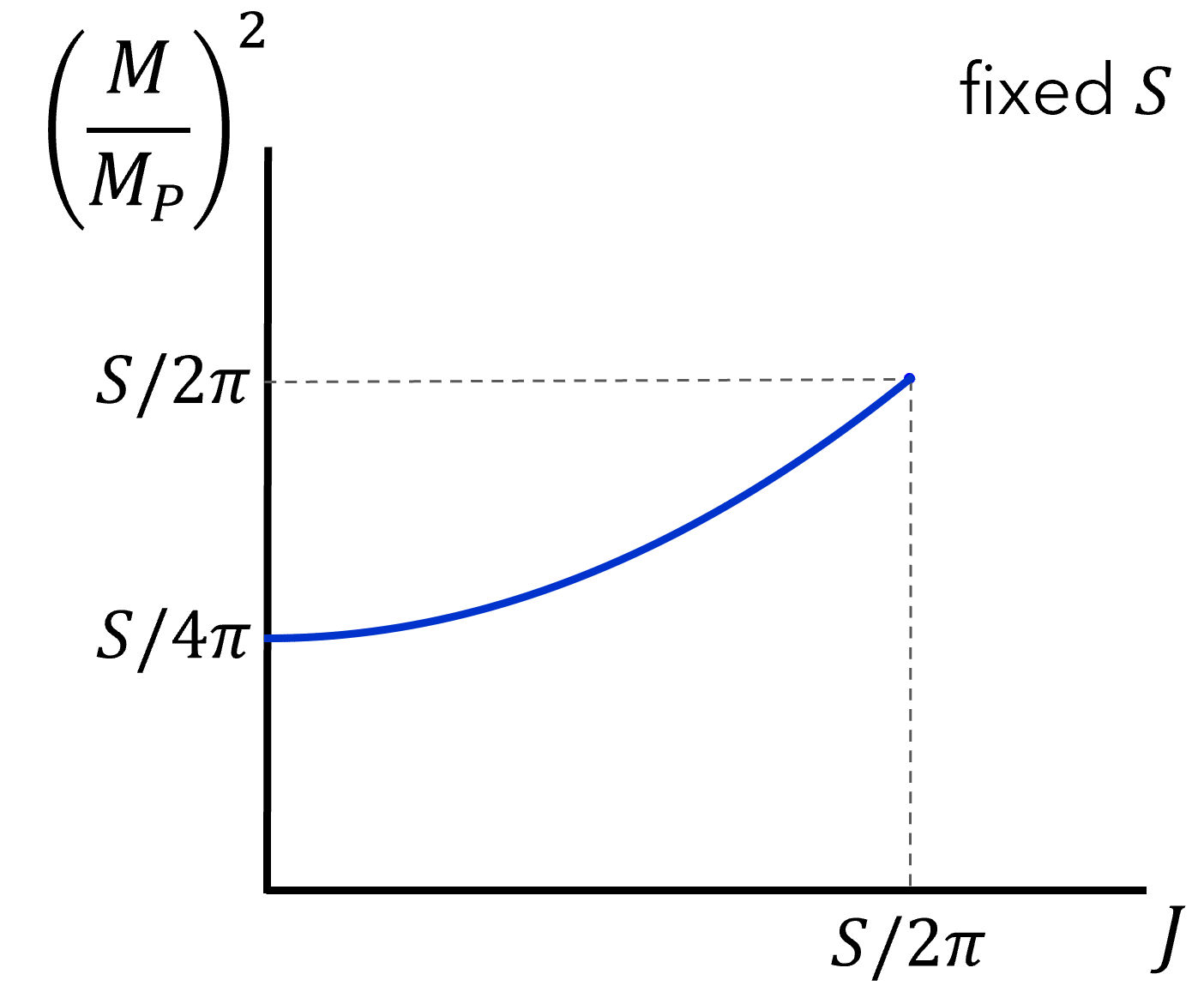}
        \caption{\small Rotating Kerr black holes in $D=4$, in a diagram of mass $M$ (in Planck units) as a function of spin $J$ for fixed entropy $S$. The spin is bounded above by the extremal limit, $J\leq S/2\pi$.}
        \label{fig:D4}
\end{figure}

The four-dimensional Kerr solution satisfies (with $G=M_P^{-2}$)
\begin{align}\label{KerrS}
   \lp\frac{M}{M_P}\rp^2 =\frac{\pi}{S}\lp \frac{S^2}{4\pi^2} + J^2\rp\,,
\end{align}
see figure~\ref{fig:D4}.
The extremal limit is reached for
\begin{align}\label{extJ}
    J=\frac{S}{2\pi}\,.
\end{align}
Values of $J$ higher than this are not relevant to us: although seemingly allowed in \eqref{KerrS}, they correspond to the entropy of the inner horizon.

The expressions \eqref{KerrS} and \eqref{extJ} are exact, but since we are not interested in $\ord{1}$ numerical factors, we will simplify the Kerr bound for fixed entropy to
the form
\begin{align}\label{kerrb}
    J \leq  S \qquad \textrm{(Kerr bound)}\,.
\end{align}
We will presently see that this is the same as the generalized Kerr bound \eqref{kerrb1} in all dimensions $D$.

\begin{figure}
        \centering
         \includegraphics[width=.65\textwidth]{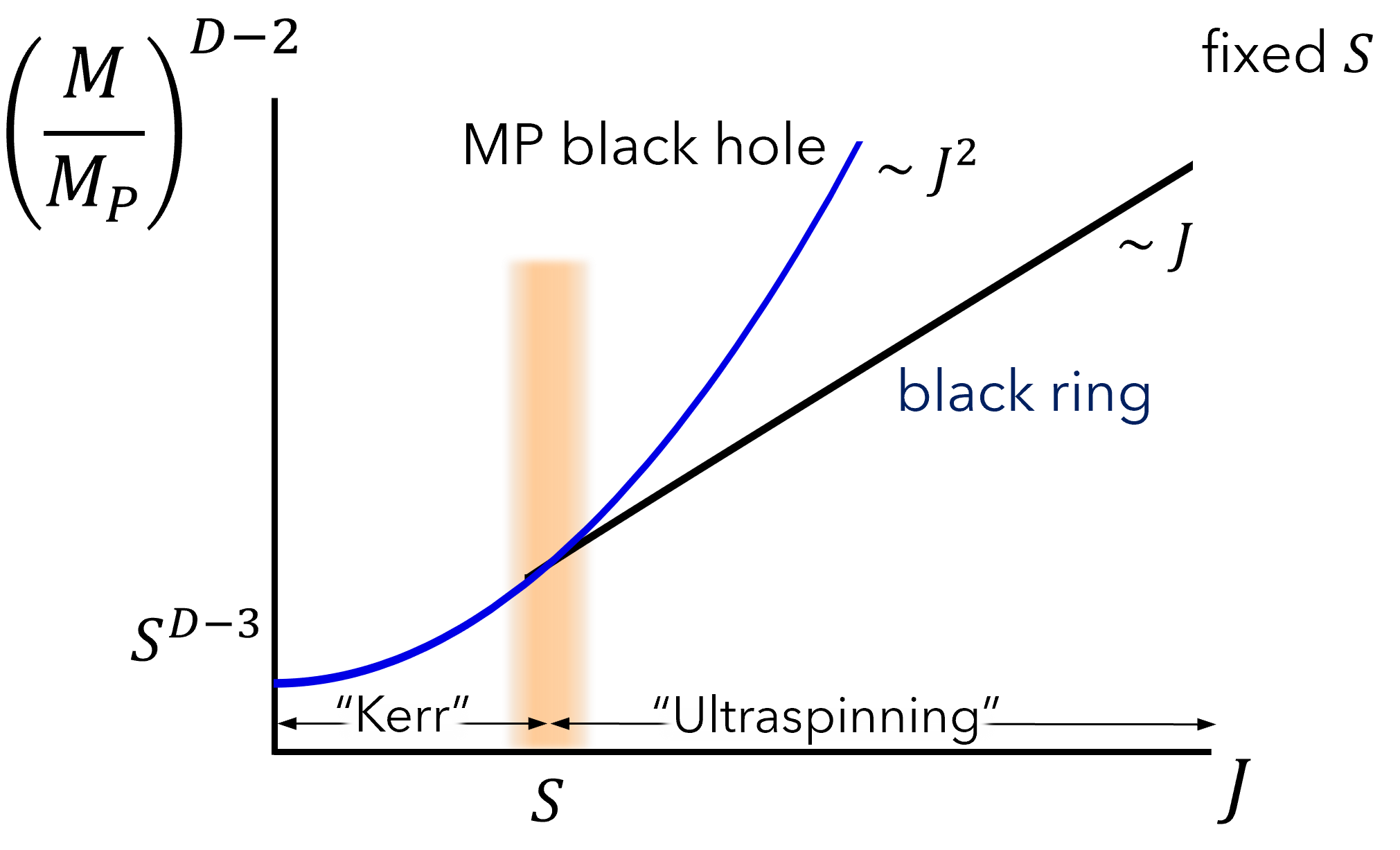}
        \caption{\small Main phases of rotating black holes in $D\geq 5$, for fixed entropy $S$ (eqs.~\eqref{MPD} and \eqref{BRD}). The diagram, plotted for the exact solutions in 5D, neglects $\ord{1}$ numerical factors but is valid for $D\geq 5$. The shaded band around $J= S$ separates the `Kerr regime' of round-shaped, stable black holes from the `ultraspinning regime' of unstable, elongated black holes. In $D=5$ the MP black hole is always stable but for $J\gtrsim S$ it is strongly distorted with large equatorial curvature. The diagram for $D=4$ (figure~\ref{fig:D4}) can be included as consisting of only the Kerr phase. [For the connoisseur: We neglect the finer structure of phases within the shaded band and in the ultraspinning range. We do not include bumpy black holes in $D\geq 6$, which interpolate between MP black holes and black rings; nor fat rings in $D=5$, which extend to all $J>S$ coinciding almost exactly with the MP curve; nor solutions with disconnected horizons.]}
        \label{fig:Dg5}
\end{figure}

For the single-spin MP black holes, for any $D$ the relation \eqref{KerrS} generalizes to 
\begin{align}\label{MPD}
   \lp\frac{M}{M_P}\rp^{D-2} = c_D\, S^{D-5}\lp \frac{S^2}{4\pi^2} + J^2\rp\,,
\end{align}
with $c_D$ a numerical factor that we give in appendix~\ref{app:hidbhs}. Now in any $D\geq 5$ these black holes can have arbitrarily large $J$ for a given $S$ (see figure~\ref{fig:Dg5}). This is the case even in $D=5$, where there is an extremal limit. However, these extremal solutions have $S=0$, so if we keep the entropy fixed we can never reach them. Extremality is, then, only asymptotically achieved when $M$ and $J$ are very large. 

Viewed this way, the five-dimensional black holes with $J> S$ are more akin to the ultraspinning solutions in $D\geq 6$, which do not have extremal limits. Indeed one can easily verify that, in $D\geq 6$, the range $J>S$ is the same as the ultraspinning regime $\ell_J>\ell_M$, where these MP black holes are dynamically unstable.
However, in $D=5$ the `bound' \eqref{kerrb} is not an extremal or stability bound (this MP black hole is dynamically stable \cite{Shibata:2010wz,Dias:2014eua,Bantilan:2019bvf}), but rather a less sharp, broader band that separates the region of low spins where the black hole is round and smooth, from the regime of higher spins where the horizon develops a highly distorted shape. In a moment we will examine this and its implications.

The other main phase of black holes consists of black rings. They are known in exact form in 5D, where they have two branches, thin and fat. Fat rings closely resemble MP black holes, only with a small hole punched in the axis, and are unstable to collapse, closing the hole and becoming MP black holes \cite{Elvang:2006dd}. We will only consider thin black rings, for which there are approximate constructions in all $D\geq 6$ \cite{Emparan:2007wm,Armas:2014bia,Dias:2014cia}. Their spin is bounded below, $J\geq S$, so they only exist in the ultraspinning regime, and in all $D\geq 5$ they very approximately follow
\begin{align}\label{BRD}
    \lp\frac{M}{M_P}\rp^{D-2} = J\, S^{D-4} \qquad (J\geq S)\,.
\end{align}
The precise numerical proportionality factor is given in appendix~\ref{app:hidbhs}.

Figure~\ref{fig:Dg5} summarizes the main black hole phases in any $D\geq 4$.  
To understand their evolution as the coupling $g$ decreases with constant $S$, we mostly need to focus on the differences between the two regimes: Kerr and ultraspinning.

\subsection{Horizon shapes and curvatures}

In the Kerr regime, black holes have a smooth, fairly rounded horizon, so their curvature is well approximated by the static value \eqref{Kbh} 
\begin{align}\label{Kstat}
     \mathcal{K}= \frac1{\ell_P^4}S^{-\frac4{D-2}}\,.
\end{align}

Neglecting any factors that are $O(1)$ in this regime, the horizon angular velocity of these black holes is
\begin{align}
    \Omega=\frac{J}{M r_H^2}\,.
\end{align}
We can assign them a moment of inertia 
\begin{align}
    I=\frac{J}{\Omega_H}=Mr_H^2\,,
\end{align}
as expected of a sphere of radius $r_H$.

In contrast, ultraspinning black holes with $J>S$ are far from round.  In $D=5$, the MP black hole gets strongly distorted as the angular momentum increases towards extremality, with the horizon flattening on the rotation plane and the curvature diverging along the equatorial rim. More explicitly, the curvatures at the rotation axis and at the equator\footnote{The precise results are in the appendix~\ref{app:hidbhs}.}
\begin{align}
    \mathcal{K}\big|_{\rm axis}&=\frac1{\ell_P^4}\frac1{(S^2+J^2)^{2/3}}\,,\label{5K0}\\
    \mathcal{K}\big|_{\rm equator}&=\frac1{\ell_P^4}\frac{(S^2+J^2)^{10/3}}{S^8}\,,\label{5Keq}
\end{align}
agree with \eqref{Kstat} when $J<S$, but begin to differ widely when $J>S$. The equatorial curvature grows arbitrarily large either in the proper extremal limit of $S\to 0$ for fixed $J$, or in the limit $J\to\infty$ for fixed $S$. Near the pole, when $J$ is large the curvature becomes very small. These effects will determine where the horizon first becomes stringy as we lower $g$ for these black holes.

The MP black holes in $D\geq 6$ never develop naked singularities for any finite $M$ and $J$, but when the spin is large their shape is highly pancaked. When $J\gg S$ the polar and equatorial curvatures are
\begin{align}\label{DK0}
    \mathcal{K}\big|_{\rm axis}&=\frac1{\ell_P^4}S^{-\frac4{D-2}}\lp\frac{J}{S}\rp^\frac{8}{D-2}\,,
\\
\label{DKeq}
    \mathcal{K}\big|_{\rm equator}&=\frac1{\ell_P^4}S^{-\frac4{D-2}}\lp\frac{J}{S}\rp^\frac{4D}{D-2}\,.
\end{align}
Both grow large with $J$, but more so along the equator. For black rings, there is a similar effect. When spun up, they become thinner and their curvature increases like
\begin{align}\label{Kring}
    \mathcal{K}=\frac1{\ell_P^4}S^{-\frac4{D-2}}\lp\frac{J}{S}\rp^\frac{4}{D-2}\,.
\end{align}
As could be expected, this is a lower curvature than for MP black holes with the same entropy and spin. 

Unlike the case of 5D MP black holes, these results for MP black holes in $D\geq 6$ and black rings in all $D\geq 5$ are of secondary importance for the correspondence, since they are overshadowed by a more dramatic phenomenon: their ultraspinning instabilities.

\subsection{Instabilities: Death by radiation and by fragmentation}

Black holes in the Kerr range of low spin are believed to be classically stable in all dimensions \cite{Emparan:2008eg}. In $D=4$ this comprises all there is. In $D=5$, the MP black hole is stable up to the extremal limit, but as we have seen, it has issues when $J>S$. 

All other black holes with $J>S$---i.e., MP black holes in $D\geq 6$, and black rings in all $D\geq 5$---suffer from instabilities.
The two main decay channels are \cite{Emparan:2003sy}: 
\begin{itemize}
\item Bar instabilities, where the fast-spinning black hole develops a bar shape on the rotation plane. The rotating black bar radiates away the `excess spin' through gravitational wave emission until it reaches the stable Kerr regime: \emph{Death by radiation}.
\item Gregory-Laflamme (GL) instabilities \cite{Gregory:1993vy}, where the pancaked black hole or thin black ring grow inhomogeneities along the horizon, until they pinch off and break up the black hole: \emph{Death by fragmentation}.
\end{itemize}
Which one dominates depends on the spin and the number of dimensions in ways that have not been fully ascertained yet, but there are general patterns. 

Bar instabilities of black holes are known to occur for lower spins than GL instabilities and likely dominate at relatively low ultraspins. The duration of a bar-shaped black hole as it radiates spin away until it becomes a stable black hole, has not been determined very precisely, but it is parametrically of the order of the horizon size $r_H$. In $D=6$ the timescale is $\sim 100\, r_H$ and in higher $D$ it is longer since gravitational wave emission is more suppressed \cite{Shibata:2010wz,Andrade:2018nsz,Andrade:2019edf}. Then, black bars possibly last long enough to enter the Goldilocks adiabaticity regime---more so for larger $D$. In contrast, any bar-shaped black holes in $D=4,5$ (however they may form) will shed off their excess spin quickly, on a time $\simeq r_H$ which is at the limit of the adiabaticity regime. 

When the spins are larger, $J\gg S$, the black holes spread much more along the plane of rotation than in the transverse dimensions. Then, the GL instability is expected to be dominant \cite{Gregory:1993vy,Harmark:2007md}. The timescale for fragmentation is parametrically of the order of the horizon size transverse to the rotation plane, $r_H$. In $D=6,7$ the time to pinch to zero is $\sim 10-100\, r_H$ \cite{Andrade:2020dgc}, and it is shorter for higher $D$. It may then seem that some pinching black holes may narrowly enter the Goldilocks regime and transition to strings before fragmenting. However, this is a chance only for moderately large spins, since the adiabaticity limit set by the absence of absorption of the dilaton wave is a timescale as long as $\ell_J$, which can be much longer than $r_H$. Therefore, save for possible limited exceptions, fragmentation decays are typically too fast to be in the Goldilocks regime.

For black rings, elastic instabilities similar to bar instabilities may dominate at moderate spins, leading to either death by radiation or by fragmentation \cite{Figueras:2015hkb}. These are, again, naturally expected to be too fast for adiabaticity.

Without attempting precision of detail, we can then draw a useful broad picture:
\begin{itemize}
\item Death by radiation of black bars will be dominant at moderately high spins. In $D\geq 6$ it may be slow enough to allow an adiabatic transition into a string state. 
\item Death by fragmentation of black pancakes and black rings will be dominant at higher spins. Pinch-off and break-up typically occur too fast to adiabatically reach a correspondence transition.
\end{itemize}

We expect that the two decay modes closely compete in some ranges of parameters (moderate-to-large spins and $D\simeq 5-7$) and initial conditions. But the end result is qualitatively the same: one or more stable black holes in the Kerr regime. Some mass is radiated away, but the main effect is the reduction of the intrinsic spin of the black holes. In the first process, this is due to classical radiation emission, and in the second one, also to the conversion of spin into orbital angular momentum of a system of several black holes. 

\subsection{Spontaneous superradiant decay near extremality}\label{subsec:srad}

In our previous discussion of the effect of Hawking decay in the correspondence, the typical black hole temperature was set by the horizon radius, which then approaches the Hagedorn temperature near the correspondence. 
This does not always hold, though, since the temperature of the Kerr black hole vanishes in the extremal limit. Nevertheless,  extremal rotating black holes still decay through non-thermal spontaneous quantum emission of superradiant modes, which leads to a fast spin-down of the black hole that takes it away from the extremal limit \cite{Page:1976df}.

The rate of this quantum superradiant emission, like the Hawking decay rate, increases as $g$ is lowered. One might then wonder if, as a consequence, extremal and near-extremal Kerr solutions will spin down so fast that they do not make it across the correspondence, but instead quickly move finitely away from extremality\footnote{Near-extremality means a deviation $\propto 1/S$ from the extremal limit, as we will see in \eqref{nearext}. ``Finitely away'' means a deviation of order one in $S$.}. 

However, this is not the case. In section~\ref{subsec:goldilocks} we saw that the relative rate of entropy loss due to Hawking emission is $T/S$. This is highest near the correspondence, where it reaches $1/(S\ell_s)$.
The rate at which the black hole spins down away from extremality (\ie the rate of change of $J/GM^2$) is of the same order, up to a factor that can be large but which remains finite in the classical limit and therefore must be smaller than $S$. Therefore the spin-down rate will be small if $S$ is sufficiently large, and the near-extremal black hole can adiabatically reach the correspondence point.  Put another way, when the correspondence is approached, the black hole, although of string-scale size, is still very large in Planck units and therefore behaves much like a semiclassical black hole. As a consequence, the effects on a large black hole (with $S\gg 1$) of phenomena like Hawking radiation and spontaneous superradiant emission are relatively very small. This forces the consideration of black holes arbitrarily close to extremality in the correspondence picture, which we will see is problematic.

Although we find these arguments robust, let us mention that many aspects of the quantum physics of near-extremal black holes may need to be reassessed following the recent work of \cite{Iliesiu:2020qvm}.

\section{Massive spinning string states}\label{sec:stringstates}

We must now discuss spinning string states that are highly degenerate, that is, there are many different states with the same mass and total spin.

The simplest construction is as solutions of the classical Nambu-Goto string, which capture many of the main properties of massive strings. Good estimates of their degeneracy can be obtained, but a proper calculation requires the quantized string, so we will also construct massive, degenerate quantum string states well approximated by classical string solutions. There are several alternative constructions using different gauge choices, and we mostly follow \cite{Blanco-Pillado:2007eit} (see also \cite{Blanco-Pillado:2007hzh}). 

In the equations in this section we retain most numerical factors and often use $\alpha'$ instead of $\ell_s$ for ease of comparison with the literature.

\subsection{Classical strings}
 We begin with the Nambu-Goto action
\begin{equation}
    I_{NG}=-\frac1{2\pi\alpha'}\int \sqrt{-\gamma}\, d\tau\, d\sigma\,,
\end{equation}
for closed strings, with $0\leq\sigma\leq \pi$ and worldsheet metric $\gamma_{mn}=\eta_{\mu\nu}\partial_m X^\mu \partial_n X^\nu$. We choose conformal gauge $\gamma_{mn}=\sqrt{-\gamma}\,\eta_{mn}$ and static gauge $X^0=2\alpha' M\,\tau$, where $M$ is the total mass of the string. With these choices, the physical velocity is orthogonal to the string, and the equations of motion are generally solved by
\begin{equation}
    X^i=X^i_R(\sigma^-)+X^i_L(\sigma^+)
\end{equation}
where $\sigma^\pm= \tau\pm \sigma$ and $i=1,\dots,D-1$ denote spatial directions.

Our gauge fixing requires that we impose the constraints
\begin{equation}\label{classconstraints}
    |\partial_\sigma X^i_R|^2=|\partial_\sigma X^i_L|^2=\alpha'^2 M^2
\end{equation}
(sum over all $i$ is implicit). This means that the energy is equally distributed among the left and right-moving modes $X^i_L$ and $X^i_R$. Other than this, the profiles for $X^i_{L,R}$ are arbitrary, and this is the origin of the large degeneracy of string states for fixed total mass $M$---infinite unless we introduce some discretization.

Since the theory is free, the probability distribution of string states will be Gaussian. This implies that typical string states will have the profile of a random walk (see \cite{Manes:2004nd} for details). Then, since the length of the string is proportional to $M$, the mean-square size of the typical string state will be given by
\begin{equation}\label{StaticStringRandomWalk}
    \langle (X^i)^2\rangle\sim (\alpha')^{3/2} M\,.
\end{equation}
If the steps of the walk have length $\sim \sqrt{\alpha'}$ and are taken at random, we expect the degeneracy of states to be\footnote{The numerical proportionality factor depends on the type of closed string (bosonic, type II, heterotic), which we do not specify. Up to this factor, the results for the entropy are precise and consistent throughout this section.}
\begin{equation}
    S\propto \sqrt{\alpha'}M\,.
\end{equation}

\subsection{Degenerate spinning string bars}

The angular momentum of the string state in the plane $(1,2)$ is
\begin{equation}\label{J12}
    J=\frac1{2\pi\alpha'}\int_0^\pi d\sigma \left( X^2\partial_\tau X^1-X^1\partial_\tau X^2\right)\,.
\end{equation}

Fixing $J$ reduces some of the freedom in choosing the oscillation profile in the rotation plane, but we want to still have a large degeneracy. What is the best way to achieve this? A general principle is to use up as little energy as possible for angular momentum by having it carried by long-wavelength excitations spread on the rotation plane. Then, a large fraction of the total energy is left to random oscillations, i.e.,  `heat'. In section~\ref{subsec:hybrids} we will encounter another realization of this strategy. 

To build our states, we put a string profile of a rigidly rotating rod in the $(1,2)$ plane. On top of this, we add a random set of oscillations in the directions transverse to the rotation plane. We will call these configurations `string bars'.
It is also easy to construct other solutions where the profile of the string in the $(1,2)$ plane is arbitrary, but, since they are less efficient at carrying the angular momentum than the rods, they will be subdominant in entropy. Nevertheless, circular profiles (`plasmid strings') are of interest and we will discuss their role in the correspondence in section~\ref{sec:dipoleplasmids}.

We do not expect that this construction of string bars strictly maximizes the entropy for given $M$ and $J$, since we are not putting any wiggliness in the rotation plane. However, this only reduces the entropy by an overall $\ord{1}$ number, outside the sensitivity of the correspondence principle. Wiggles on the rotation plane are technically problematic since they lead to self-intersections in the rotation plane. When interactions are turned on, these can break up the string profile in the plane. In four spacetime dimensions this is unavoidable---in order to have independent oscillations, our construction requires at least four space dimensions---and is in line with the shorter-lived nature of spinning states in $D=4$. We will not attempt to improve on this issue. The construction described above is simple, explicit, and can readily be extended to the quantum string.

Let us then separate the oscillations in the $(1,2)$ plane from all other directions,
\begin{equation}
    X^i=(x^1,x^2,X^k)=(\mathbf{x},X^k)\,,\qquad k=3,\dots, D-1\,.
\end{equation}
To carry total angular momentum $J$, we choose
\begin{equation}\label{rods}
    \mathbf{x}_L= \sqrt{\frac{\alpha' J}2}\, (\sin 2\sigma^+,\cos 2\sigma^+)\,,\qquad 
    \mathbf{x}_R= \sqrt{\frac{\alpha' J}2}\, (\sin 2\sigma^-,\cos 2\sigma^-)\,. 
\end{equation}
The left and right profiles are circles, but when added they describe a finite segment, or rod, rotating on the plane. 

The oscillations in the transverse plane are now constrained by \eqref{classconstraints} to satisfy
\begin{equation}\label{constr2}
   \frac1{\alpha'}|\partial_\sigma X^k_{L,R}|^2= \alpha'M^2-2J\,,
\end{equation}
 (sum over $k$ is implicit) and therefore the total spin is bounded above,
\begin{equation}\label{exactregge}
    J\leq J_\text{Regge}=\frac{\alpha'}2 M^2\,.
\end{equation}
This is the Regge bound for closed strings\footnote{It differs from the bound for open strings by a factor of $2$.}. When it is saturated we recover the well-known rotating string rods with maximum angular momentum.
They are not degenerate since the constraints forbid any oscillation in the directions $X^k$ transverse to the rod. However, we can also obtain string states with strictly sub-Regge angular momenta, $J<J_\text{Regge}$,
which can be pictured as shorter, wiggly rotating rods---string bars---with a radius equal to
\begin{equation}\label{halfbar}
    R = \sqrt{2\alpha' J}\,.
\end{equation}

Given \eqref{constr2}, we expect that the degeneracies from choosing $X^k_{L,R}$ add up to a string bar entropy
\begin{equation}\label{barS}
    S \sim \sqrt{\alpha'M^2-2J}\,.
\end{equation}

\subsection{Quantum string bars}

The construction of quantum states for the previous configurations is straightforward when using a physical gauge such as light-cone gauge $X^+=(X^0+X^{D-1})/\sqrt{2}= 2\alpha' p_+ \tau$, where now $D$ is the critical dimension, i.e., 26 or 10 for bosonic or supersymmetric strings, respectively. This is a standard construction, so we omit details. 

The mass of the states is given by
\begin{equation}\label{mass2}
    \alpha' M^2 |\Psi\rangle = 4(N_L+N_R)|\Psi\rangle\,,
\end{equation}
where $N_{L,R}$ are the left and right levels of physical oscillators, and we must impose the level-matching constraint $N_L=N_R$ as a condition on physical states. For simplicity, since we will only consider highly excited states, we neglect the mass shift responsible for the tachyon mass in the bosonic string.

The angular momentum on the $(1,2)$ plane is
\begin{equation}
    J=-i\sum_{n=1}^\infty \frac1{n}\left(
    \alpha_{-n}^2\alpha_n^1-\alpha^1_{-n} \alpha^2_n 
    +\tilde\alpha_{-n}^2\tilde\alpha_n^1-\tilde\alpha^1_{-n} \tilde\alpha^2_n 
    \right)\,.
\end{equation}

In the states that we construct, $|\Psi\rangle =|\Psi\rangle_L\otimes |\tilde\Psi\rangle_R$,
the left and right-moving components have the same factorized structure,
\begin{equation}\label{12perp}
    |\Psi\rangle_L =|\psi\rangle_L\otimes |N^\perp_L\rangle_L\,.
\end{equation}
The first factor is a state of oscillators in the plane $(1,2)$,
\begin{equation}
    |\psi\rangle_L = (\alpha_{-1}^1+i\alpha_{-1}^2)^{J/2} |0\rangle_L\,,
\end{equation}
which is an eigenstate of the left-moving angular momentum with eigenvalue $J/2$.
The  second factor, $|N^\perp_L\rangle_L$, denotes an eigenstate of the left level operator for oscillators in the directions orthogonal to the plane $(1,2)$,
\begin{equation}
    N^\perp_L=\sum_{k=3}^{D-2}\sum_{n=1}^\infty \alpha_{-n}^k \alpha_n^k\,.
\end{equation}
It is then straightforward to verify that the total left level is
\begin{equation}
    N_L|\Psi\rangle_L = \left(\frac{J}2+N^\perp_L\right)|\Psi\rangle_L\,.
\end{equation}

With the right-moving sector similarly built, the state $|\Psi\rangle$ is an eigenstate of the mass and angular momentum. The level-matching condition requires
\begin{equation}
    J+2N^\perp_L =J+2 N^\perp_R=\frac{\alpha'}2 M^2\,,
\end{equation}
where the last equality follows from \eqref{mass2}. The maximum angular momentum saturating the Regge bound is obtained when $N^\perp_L =N^\perp_R=0$.

\begin{figure}
        \centering
         \includegraphics[width=.95\textwidth]{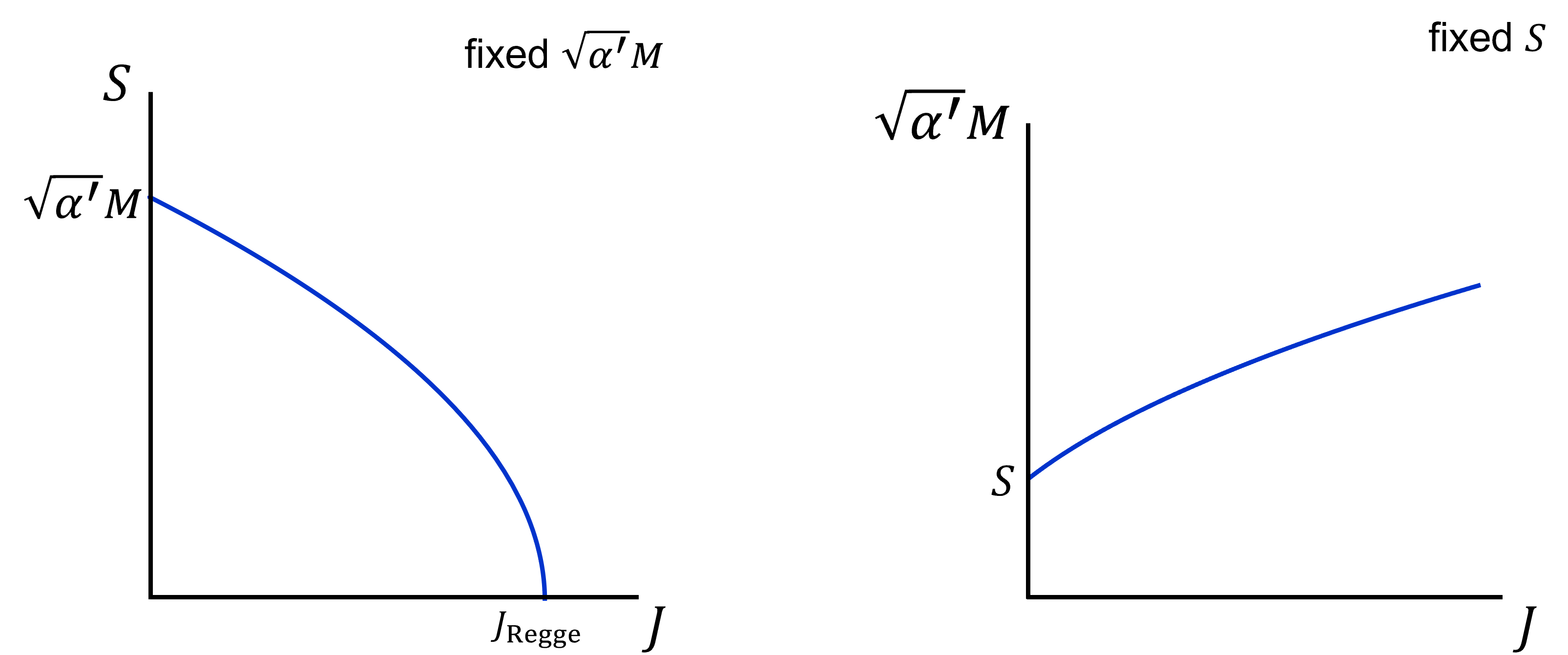}
        \caption{\small String bars. Left: $S(J)$ for fixed mass in string units \eqref{barS}. Right: $M(J)$ for fixed entropy \eqref{stringM}. We expect that the string bar states constructed in this section are close to the maximum entropy for given $M$ and $J$, or equivalently, minimum mass for given $S$ and $J$. For small but nonzero $J$, the construction misses details of typical states, see section~\ref{subsec:stringshapes}.}
        \label{fig:stringbars}
\end{figure}

The degeneracy of these states comes from the oscillators in the orthogonal directions. A standard calculation of the type in \cite{Russo:1994ev} yields the leading order result
\begin{align}\label{stringSJ}
    S &\sim \sqrt{N^\perp_L} + \sqrt{N^\perp_R}\nonumber\\
    &= \sqrt{\alpha'M^2-2J}\,.
\end{align}
 We plot it in figure~\ref{fig:stringbars}~(left). This justifies our earlier expectation \eqref{barS}.
As we did for black holes (cf.~\eqref{MPD}), it is useful to give the mass of a string bar as a function of spin for fixed entropy,
\begin{align}\label{stringM}
    \sqrt{\alpha'}M= \sqrt{S^2+2J}\,,
\end{align}
which we plot in figure~\ref{fig:stringbars}~(right). Note that for fixed $S$ there is no upper bound on $J$. The reason is the same as we explained in section~\ref{subsubsec:fixedS} for $D=5$ MP black holes.

\subsection{Sizes, shapes and degeneracies}
\label{subsec:stringshapes}

Let us measure the size of the states in the plane $(1,2)$ to compare it to the classical string configurations. For this, we compute the expectation value of the square radius
\begin{equation}
    R^2 =(X^1)^2 + (X^2)^2 =\alpha' \sum_{j=1,2}\sum_{n=1}^\infty \frac1{n^2}\left( \alpha_{-n}^j\alpha_n^j +\tilde\alpha_{-n}^j\tilde\alpha_n^j\right) +\alpha'\sum_{n=1}^\infty \frac1n\,.
\end{equation}
The last divergent term arises from operator reordering. Since it is state-independent it is customarily subtracted away \cite{Karliner:1988hd}. After appropriately normalizing $|\Psi\rangle$, one finds
\begin{equation}\label{msR}
    \langle R^2\rangle =\langle\Psi| R^2 |\Psi\rangle =  \alpha' J\,.
\end{equation}
Therefore, the rms size of a quantum string bar is $1/\sqrt{2}$ times smaller than the half-length of the classical string bar \eqref{halfbar}. This is expected: the average position (center of mass) of the rotating bar is at the origin of coordinates, while its quadratic size average $\langle R^2\rangle$ is $1/2$ of the square of its radius.\footnote{`More `semiclassical' string rods, which are not mass eigenstates but have the same length and shape as the classical solution, can also be constructed \cite{Blanco-Pillado:2007hzh}.}

We expect the size in the transverse directions to be that of a random walk of $N^\perp$ steps,
\begin{align}\label{RotatingStringRandomWalk}
     \langle (X^k)^2\rangle\sim \alpha'\sqrt{\alpha' M^2-2J}\,.
\end{align}
Then, states approaching the Regge bound \eqref{exactregge} (but still highly degenerate) are longer in the rotation plane than in the transverse directions. 

Our construction of string bars places all the random oscillations away from the rotation plane, while a typical state is expected to random-walk in all directions. This should especially affect the properties of typical states with $J$ smaller than $\ord{M/M_s}^2$. In particular, the dependence of $S$ on $J$ at small $J$ should differ from \eqref{stringM}.

The proper description of typical states follows a statistical approach where we consider the ensemble of string states with a given mass and spin. 
This study was initiated in \cite{Russo:1994ev} where the number of states of a single rotating string was computed. The entropy of slowly and highly rotating strings is given by~\cite{Russo:1994ev,Matsuo:2009sx}\footnote{Here we give the results for closed strings but the calculations in~\cite{Russo:1994ev,Matsuo:2009sx} were done for open strings. Also, in $J=O(M^2)$, $J$ should not be too close to $M^2$ for the approximation to be valid.}
\begin{equation}
\label{Sstring}
\setlength{\tabcolsep}{20pt}
\renewcommand{\arraystretch}{1.3}
S\sim \left\{\begin{array}{ll} 
     (\sqrt{\alpha'} M-c J^2) &\quad \text{for} \quad J\ll \alpha' M^2\,, \\
    \sqrt{\alpha' M^2-2J} &\quad \text{for} \quad J=\mathcal{O}(\alpha'M^2) \,,
    \end{array}\right.
\end{equation}
where the precise value of $c$ can be found in~\cite{Matsuo:2009sx}. The sizes of single-string configurations can also be computed from a statistical approach. In \cite{Mitchell:1987hr, Mitchell:1987th} it was found that the size of static strings agrees with the random-walk estimate in \eqref{StaticStringRandomWalk}~\cite{Manes:2004nd}. In a companion paper~\cite{CEPT2} we compute the size of rotating strings; in particular we find agreement with the random-walk expectation~\eqref{RotatingStringRandomWalk}.

\subsection{Interactions and decay}\label{sec:interdec}

When the string self-interactions are turned on, we expect that the random-walk ball component of these states becomes more compact due to gravitational self-attraction in the manner explained in section~\ref{subsec:sizes}, and also that it emits a thermal spectrum of massless particles, at a rate proportional to the entropy as in \eqref{Gammast},
\begin{align}\label{ballrate}
    \Gamma_{\rm ball}\sim g^2 S\ell_s^{-1}\,.
\end{align}

The string will also radiate through its motion in the  $(1,2)$ rotation plane. Since we are always in a weak-coupling regime, this emission can be treated separately from the thermal radiation from the ball. Then we can estimate the decay rates using those of string states with simple profiles as in \cite{Iengo:2006gm}. For instance, the decay rate of a closed-string rod of length $2R$ is
\begin{align}\label{rodrate}
    \Gamma_{\rm rod}\sim g^2 \frac{2R}{M}\ell_s^{-3}\sim g^2 \frac{\sqrt{J}}{M} \ell_s^{-2}\leq g^2 \ell_s^{-1}\,,
\end{align}
where we have used \eqref{halfbar}. The precise dependence on $J$ and $M$ is not important for us; it suffices to know the maximum rate, which is attained by the Regge rods.

Whenever we have a degeneracy $S\gg 1$, these decays are much slower than the thermal rate \eqref{ballrate}. The reason is that the string has a much higher chance of intersecting itself when wrapped into a ball than when spread out in a smooth profile. This will be corrected as the string coupling is increased: the ball will become more compact and radiation will take longer to escape; furthermore, a string with a varying quadrupole profile will begin to radiate more like a gravitational antenna. Still, it is easy to verify that the gravitational wave radiation rate from an ellipsoidal bar \cite{Andrade:2019edf}, when extrapolated to the correspondence point, remains much smaller than \eqref{ballrate}, and even more so for higher $D$.

The main conclusion is that, if the rate of change of $g$ is within the Goldilocks range \eqref{goldilocks}, all the string states that we have considered will reach the correspondence point keeping approximately the same overall profile in the rotation plane as at zero coupling. In the transverse directions, they will shrink from random-walk size down to string-scale length, like non-spinning string balls do.

\subsection{Black hole-string hybrids}\label{subsec:hybrids}

\begin{figure}
        \centering
         \includegraphics[width=.5\textwidth]{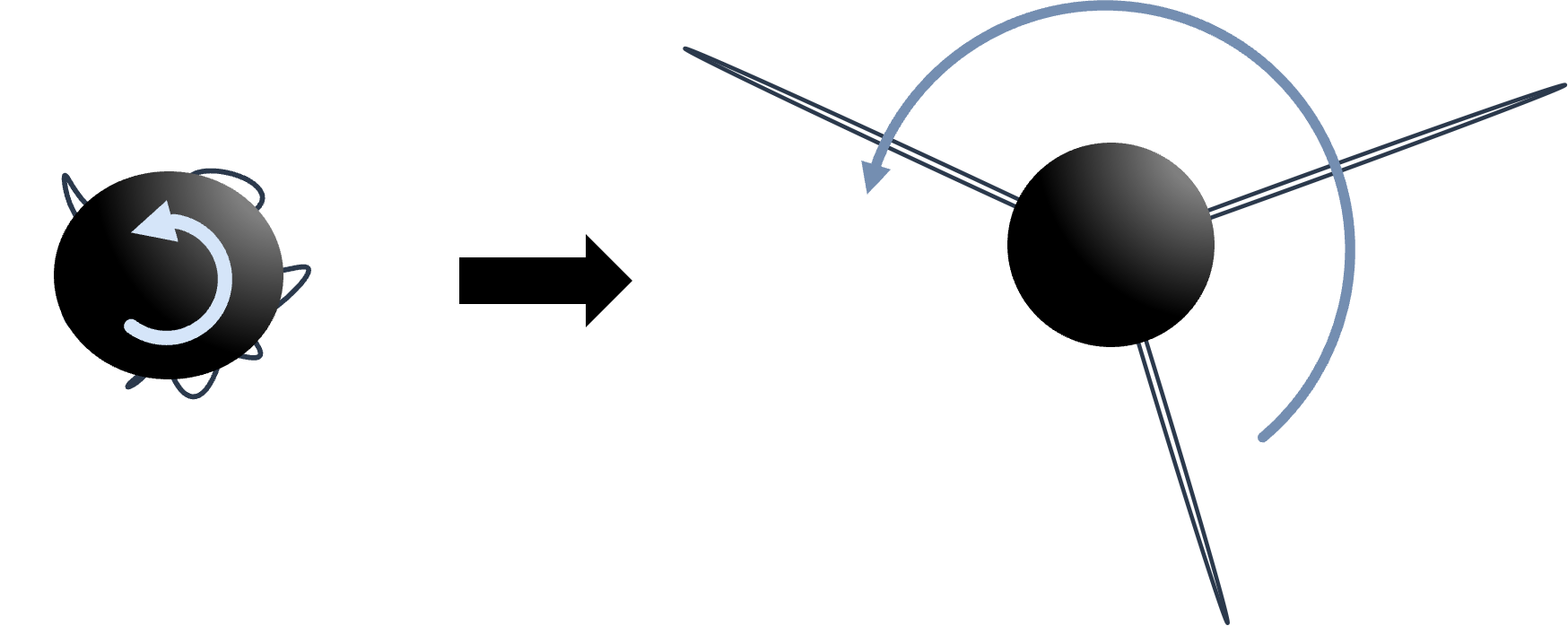}
        \caption{\small Black hole-string hybrids. Loops of string attached to a spinning black hole grow by superradiantly extracting the rotational energy of the black hole, until they become maximally spinning string rods that carry most of the initial angular momentum. After a long time, the spin is radiated away and the string is absorbed by the black hole. The simplest hybrid is a combination of a slow (Kerr-regime) black hole and a long Regge string rod.}
        \label{fig:hybrids}
\end{figure}

Configurations that combine black holes and strings will also feature in the correspondence. They provide a realization of the idea that, in order to maximize the entropy of a system with a given mass and spin, it pays to divide it into two components: a compact, highly entropic object, and a subsystem spread out on the rotation plane that carries angular momentum with as little energy as possible. For the former, we take a slowly spinning black hole (in the Kerr regime), and for the latter, a Regge-saturating string rod (or possibly more than one) sticking out of the black hole (see figure~\ref{fig:hybrids}). 

Such configurations have been studied in four dimensions with cosmic strings described by the classical Nambu-Goto action \cite{Frolov:1996xw,Snajdr:2002aa,Kinoshita:2016lqd,Igata:2018kry,Xing:2020ecz, Deng:2023cwh}. When these strings rotate on a single plane, many of their properties are independent of the number of dimensions, so we can easily adapt the conclusions of these works to our setup. We must only bear in mind the suppression of gravitational wave emission as $D$ grows, and that in our case the probability of string reconnection is $g\ll 1$, unlike $g\simeq 1$ for field-theoretic cosmic strings.

The most salient properties are readily summarized (we follow \cite{Xing:2020ecz,Deng:2023cwh}). A loop of string attached to a spinning black hole will grow by superradiant extraction of the rotational energy of the black hole in a manner computed in \cite{Xing:2020ecz}. If the length $L$ of the string is initially comparable to the black hole radius $r_H$, afterwards it increases like
\begin{align}\label{Lgrow}
    L=r_H\sqrt{1+\Omega_{\rm BH}t}\,,
\end{align}
where $\Omega_{BH}$ is the angular velocity of the horizon (neglecting numerical factors, this should be valid in any $D\geq 4$). The string stops stretching when it has extracted the spin $J$ of the initial black hole and becomes a long, Regge-like rotating rod. At that moment,
\begin{align}
    L_\mathrm{max}=\sqrt{J}\ell_s\,.
\end{align}
If $J\gg 1$, this is longer than the size of a black hole near the correspondence. Thus, an ultraspinning black hole can plausibly evolve into such a hybrid. Furthermore, since $\Omega_{BH}$ never exceeds its maximum value $\sim 1/r_H$, which is attained around $J=S$, the growth rate from \eqref{Lgrow} is always slow enough to not disturb the adiabatic evolution across the correspondence.

We will see that hybrids, as a possible evolution of ultraspinning black holes, are most important for the correspondence in $D=5$. In this case, we can easily verify that the hybrid is more entropic than the MP black hole. Regarding it as consisting of a Kerr-regime black hole and a Regge rod, the mass of the hybrid is the sum of its components, so
\begin{align}
    M=M_P\left( S^{2/3}+g^{2/3}\sqrt{J}\right)
\end{align}
and therefore
\begin{align}
    S^2=\left(\frac{M}{M_P}\right)^3\left( 1-g^{2/3}\frac{\sqrt{J}}{M/M_P}\right)^3\,.
\end{align}
We now compare this to the entropy of the 5D MP black hole \eqref{MPD},
\begin{align}
    S^2=\left(\frac{M}{M_P}\right)^3\left( 1-\frac{J^2}{(M/M_P)^3}\right)\,.
\end{align}
If the spin is near the extremal value, $J^2\simeq (M/M_P)^3$, the entropy of the MP black hole is much smaller than that of the hybrid, even more so since $g\ll 1$. If we make the comparison for fixed $S$ and $J$, then the hybrid has a lower mass, so the MP black hole can decay into it.

In turn, hybrids are not stationary systems but decay through two effects. One is the emission of gravitational waves by the spinning string, which leads to spin down. We have already seen in \eqref{rodrate} that this is slow enough to preserve the Goldilocks adiabaticity window. The other effect is the friction of the rotating string against the slower black hole horizon. This dissipation makes the string gradually fall into the black hole, at a rate  \cite{Xing:2020ecz}
\begin{align}
    \Gamma_\mathrm{fr}=\frac{r_H^2}{L^3}\,.
\end{align}
Since $L$ is the largest length in the system, this is again slow enough for adiabatic evolution. The reconnections that make the string emit long loops, which were important in \cite{Xing:2020ecz}, are negligible near the correspondence, where the coupling $g$ is very small.

Through the combined effects of wave emission and friction, a hybrid with a large total spin will, in the long run, radiate away most of its initial angular momentum and leave behind a stable, Kerr-regime black hole.\footnote{During this process the increase of entropy is at most of order $\ord{1}$. }
Nonetheless, as argued above, these decay processes are slow enough that hybrids can satisfy the Goldilocks condition \eqref{goldilocks}. As such, hybrids can survive the adiabatic evolution up to the correspondence point where they smoothly match to string bars. This is because the central, slowly spinning black-hole part of a hybrid can transition into a string ball. Combined with the string rod component, this results in an elongated stringy configuration whose transverse oscillations are localised around its midpoint.

It is interesting that in Susskind's picture of the correspondence in \cite{Susskind:1993ws}, the black hole is a sort of hybrid: strings attached to the black hole exist within a stretched horizon extending a distance $\ell_s$ away from the event horizon. Our hybrids can be regarded as a semiclassical extension of these strings, as rods whose large angular momentum makes them reach out far beyond the stretched horizon. It is suggestive to think that the short strings within the stretched horizon atmosphere of a rotating black hole grow large and semiclassical through superradiant transfer of the spin of the black hole to the strings. One might say that spin can make manifest the presence of a stringy atmosphere around black holes.

\section{Correspondence for rotating objects}
\label{sec:Correspondence}

We have now gathered all we need for establishing the correspondence between rotating black holes and strings. The presence of rotation introduces differences depending on whether the direction of the correspondence is \emph{string$\to$black hole} or \emph{black hole$\to$string}, and on the dimensionality of spacetime. These dependencies also occur when including the self-gravitation of strings \cite{Horowitz:1997jc,Damour:1999aw,Chen:2021dsw}, as we saw in section~\ref{subsec:sizes}, but with rotation they appear even before including such effects.

\subsection{Strings to black holes}\label{subsec:StBH}

This correspondence was illustrated in figure~\ref{fig:StrBH}. Now we provide a more detailed discussion.

\paragraph{Slow string balls: $J<S$.} The evolution of these strings as $g$ grows is essentially the same as in the spinless case, since the angular momentum only modifies physical magnitudes by $\ord{1}$ factors. These states are fairly round-shaped and, to a first approximation, when $g^2=1/S$ they evolve into Kerr-regime, stable black holes. In more detail:
\begin{itemize}
    \item The entropy of the black hole depends on $J^2$ (not $J$). This is also observed for string states with $J<S$ in \eqref{Sstring}. Further detail of the $J$ dependence is beyond the accuracy of the correspondence.
    \item In $D=4$ the transition of strings with $J$ finely tuned to the (near-)extremal Kerr black hole parameters may not be adiabatic. We will discuss this further in section~\ref{subsec:BHSt4D}. 
    \item Self-gravitation of the string introduces important dependence on $D$ (section~\ref{subsec:sizes}). Much of what has been found for spinless states carries over to this regime of slow spins. Self-gravity can also be relevant for the features discussed in the previous two points. 
\end{itemize}

\paragraph{Fast string bars: $J>S$.} We identified string bars, or wiggly rods, as the most relevant (most entropic) states in this regime (plasmids will be discussed in section~\ref{sec:dipoleplasmids}). 
As we mentioned in the introduction, they pose an immediate apparent puzzle in $D=4$, since there are no stationary black holes in this range. 

The authors of \cite{Bardeen:1999px} observed that the length of these strings is at least $R\gtrsim J/M\gtrsim  \ell_s J/S$, which is larger than the size $\ell_s$ expected of a black hole at the correspondence coupling. Then they argued that these strings would become black holes only when the coupling becomes large enough that all of the string is within its Schwarzschild-Kerr radius, $R\lesssim GM$. This happens when
\begin{align}
    g^2 = \frac{R}{\ell_s S}\gg \frac1{S}\,,
\end{align}
at which point the string would collapse to form a Kerr black hole close to extremality.
Observe, however, that this collapse would not be adiabatic, since the entropy would jump discontinuously,
\begin{align}
    S_\mathrm{BH}=J> S_\mathrm{st}\,.
\end{align}
This jump does not immediately imply a contradiction, but only a breakdown of the picture of a smooth correspondence. One might also think that the puzzle could disappear in $D>4$, where there do exist stationary ultraspinning black holes with $J>S$. 
However, while the picture of non-adiabatic collapse may be realized for string states of very low degeneracy very close to the Regge bound, a more compelling evolution is possible for the highly degenerate states of string bars. 

The string bar is a rod with a random wiggly structure, and as the coupling $g$ grows, 
we expect that the wiggly-ball part of it will become more compact, whereas the rod does not get any significantly shorter.
Therefore, we expect that the string, rather than suddenly collapsing as a whole to form a MP black hole at a high value of the coupling, will instead form a non-stationary black bar or a black-hole-string hybrid when $g^2=1/S$. Observe that when the coupling is slightly above the correspondence value, the hybrid and the bar have more entropy than the wiggly rod because their entropy is that of a black hole, so their formation seems indeed favored\footnote{If we fix the entropy, they have less total mass, so they are viable decay states.}. Which of the two occurs probably depends on how large the spin is and very likely on the number of dimensions $D$.

In $D=4$, the long range of the gravitational interaction, by way of the `hoop conjecture', disfavors long black bars (for the same reason that there are no black strings in $D=4$) so bars could only be short-lived and short, and would seem to be an option only slightly above the Kerr bound. The most likely evolution of the wiggly string bar is that its random-walk-ball part collapses into a Kerr black hole, with the rod spiking out of it. This would be a hybrid. 

In $D\geq 5$ there is no hoop conjecture to ban black bars, which can be long-lived objects that would form from the collapse along the length of the wiggly bar. The shorter-range interaction will actually be less efficient at pulling the mass toward the center of the state to form a hybrid. Thus, rotating black bars appear as a more likely option in $D\geq 5$. 

To summarize the evidence, it seems plausible that fast-spinning string bars evolve into
\begin{itemize}
    \item black hole-string hybrids in $D=4$,  
    \item black bars in $D\geq 5$. 
\end{itemize}
Bear in mind, though, that our considerations above are rough and the actual transitions may be more complex, possibly involving both kinds of objects, \eg a black bar may evolve into a hybrid.

This picture is more nuanced than the delayed-collapse model suggested in \cite{Bardeen:1999px} and in particular it allows for adiabatic correspondence.
In the formation of a hybrid, the string-ball part carrying most of the entropy will form a black hole in an adiabatic way, providing a smooth transition in $D=4$. The collapse into a black bar in $D\geq 5$ is not controlled by the length of the bar but by its transverse size. Since this transition is, at each point along the length of the bar, akin to that of \emph{string ball$\to$black hole} in $D-1$ dimensions, we can expect it to be adiabatic at least in $D=5,6$, while in higher dimensions self-gravitation effects must be clarified to discern the picture.

\subsection{Black holes to strings: $D=4$}\label{subsec:BHSt4D}

The main features of the correspondence were illustrated in figure~\ref{fig:BHStr46} (left).

We imagine starting from a Kerr black hole at a large value of the coupling, which we then reduce until the curvature grows so large that stringy effects become important. The curvature near a Kerr black hole is almost uniform on the horizon, so when $g^2=1/S$ the black hole as a whole will morph into a string state, presumably a slowly rotating string ball with $J<S$.

Since the angular momentum $J$ of the black hole cannot be larger than $S$, the mass changes parametrically smoothly across this transition as it does in the absence of spin \cite{Matsuo:2009sx}. The black hole rotates at a larger angular velocity than a rotating string ball, but when self-gravity shrinks the latter to a size $\sim \ell_s$, its angular velocity increases, ballerina-like, to match that of the black hole.

We must also check the continuity of the temperature across the transition.
The black hole temperature \eqref{TOmSJ} at the correspondence point, where $\ell_P=\ell_s/\sqrt{S}$, is 
\begin{align}
    T_H=\frac1{\ell_s}\frac{S^2-4\pi^2 J^2}{S\sqrt{S^2+4\pi^2 J^2}}\,
\end{align}
(neglecting overall numerical factors). For black holes that are sufficiently far from the extremal limit \eqref{extJ}, with
\begin{align}
    S-2\pi J =\ord{S}\,,
\end{align}
this is the same as the Hagedorn temperature $T\sim 1/\ell_s$ of the string ball, so the transition can proceed smoothly. However, near extremality, with
\begin{align}\label{nearext}
    S-2\pi J \ll S\,,
\end{align}
the temperature of the black hole
\begin{align}
    T_H=\frac1{\ell_s}\frac{S-2\pi J}{S}
\end{align}
is much smaller than the string temperature and the transition would seem discontinuous.\footnote{This was observed in \cite{Matsuo:2009sx} but henceforth our discussion differs.} As we discussed in section~\ref{subsec:srad}, the spin-down due to non-thermal superradiant emission is not fast enough to prevent the near-extremal black hole from reaching this point. 

\paragraph{Choking the throat.} 

Near-extremal Kerr black holes have long throats \cite{Bardeen:1999px} with almost constant curvature---the polar angular distortion is no more than an $\ord{1}$ effect. So when the correspondence coupling is reached, all of the throat becomes a string state.

As we have seen, there is no problem in finding a string ball with parametrically the same mass, entropy, spin and rounded shape as a near-extremal black hole. It is then possible that the throat chokes by forming a string ball, but the fact that the temperature would change abruptly indicates that the nature (although not the number) of the degrees of freedom of the system changes across the transition. So this would not be an adiabatic change.
It is an interesting problem whether, by increasing the self-gravitation of a spinning string ball, a throat develops or not.

The question of which string degrees of freedom describe the near-extremal Kerr throat has remained mysterious \cite{Guica:2008mu,Compere:2012jk}. All we can say is that at the correspondence point, they do not seem to be those of conventional string balls. If the recent studies of charged black hole throats in \cite{Iliesiu:2020qvm} extend also to the extremal Kerr solution, then large quantum effects can significantly affect the picture.

\subsection{Black holes to strings: $D \geq 6$}\label{subsec:BHSt6D}

This correspondence was illustrated in figure~\ref{fig:BHStr46} (right).

\paragraph{Kerr black holes: $J<S$.} These black holes evolve in a simple way: they become spinning string balls at the correspondence without adding any novelty to the spinless transition picture. Since there is no extremal limit for any $J$, the difficulties that we found in $D=4$ are absent.

\paragraph{Ultraspinning black holes: $J>S$.} The evolution of these black holes is dominated by their instabilities, which will be triggered, if not by any other means, by quantum effects either in the form of Hawking emission or as higher-curvature corrections. We argued that at moderately high spins, bar instabilities are the most likely decay, while at higher spins fragmentation should be dominant.

Bar instabilities drive the black hole, through the emission of gravitational radiation, to a smaller ratio $J/S$. If the radiation rate is significantly slower than $1/\ell_s$, (which is more likely when $D$ is larger), then the rotating black bar will smoothly proceed through the correspondence to become a rotating stringy bar. If instead, the spin loss is fast, $\sim 1/r_H$, the black hole will radiate away the excess spin, enter the stable Kerr regime and then transition into a string ball. 

Fragmentation instabilities are very quick, so ultraspinning black holes and black rings are unlikely to remain in the Goldilocks adiabaticity window. Instead, they will first fragment into several Kerr-like black holes, which, when reaching the correspondence, will transit to string balls. If the number of fragments is not extremely large (less than a power of $S$) then their individual entropies will be $\ord{1}$ fractions of the initial $S$ and these transitions will occur at a coupling parametrically the same as \eqref{gcorrS}.  

Notice that in the fragmentation scenario, the final state of the evolution is not a single-string state (plus some radiation), but rather one with multiple massive strings. This aligns with the early proposal in \cite{Emparan:2004wy} that thin neutral black rings should be understood in terms of multi-string states. 

Hybrids do not seem necessary but could play a role in some parameter ranges. 
To see how, we turn to the Kretschmann scalar along the equator of an ultraspinning black hole, computed in \eqref{DKeq}. It reaches the string scale when
\begin{align}\label{gusbh}
    g^2=\left(\frac{J}{S}\right)^D\frac1{S}\,,
\end{align}
which is a larger value than the correspondence coupling for Kerr black holes. If an ultraspinning black hole somehow avoided death by radiation or fragmentation and reached this point, we would expect it to become stringy near the equator and develop a hybrid form. Given the quickness of ultraspinning instabilities, this may be an unlikely route, but we will see that for five-dimensional black holes hybridization must be reckoned with.

For a thin black ring, the curvature \eqref{Kring} on the horizon reaches the string scale when
\begin{align}\label{gring}
    g^2=\frac{J}{S^2}=\frac{R}{\ell_s}\,\frac1{S}\,,
\end{align}
which is again larger than $1/S$. If the black ring managed to reach this coupling intact, we would expect it to become a ring-shaped string state very unstable to breaking apart into multiple string balls, \ie not a very different outcome than if the black ring had fragmented in the first place. The existence of this instability can indeed be verified using the formalism in \cite{Horowitz:1997jc} for self-gravitating string states \cite{ESGT}.

\subsection{Black holes to strings: $D=5$}\label{subsec:BHSt5D}
\begin{figure}
        \centering
         \includegraphics[width=.65\textwidth]{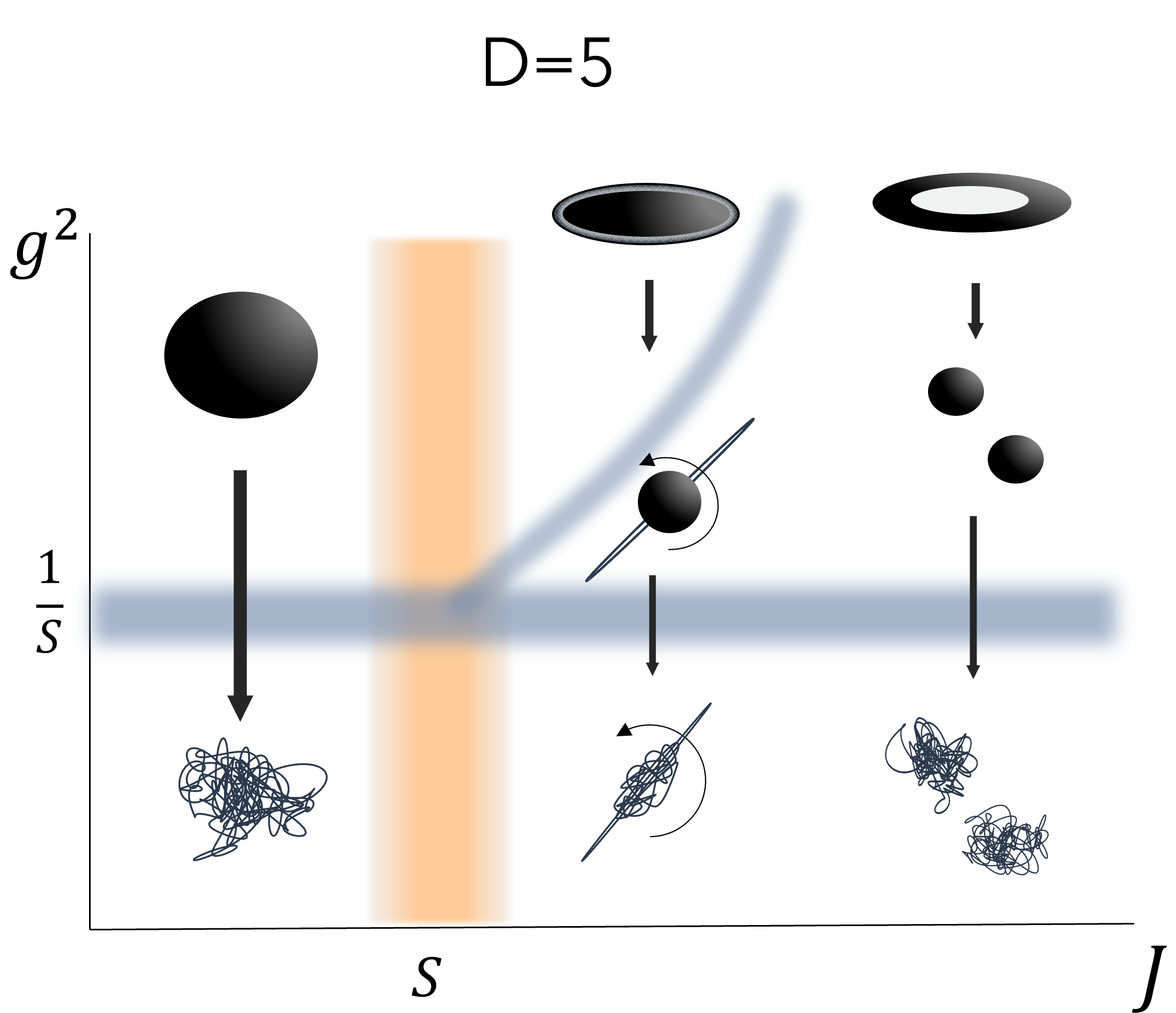}
        \caption{\small Correspondence from black holes to string states in $D=5$. For Kerr-like black holes with $J<S$ the picture is the same as in $D\geq 6$. MP black holes with $J>S$ are dynamically stable, but they develop large curvatures near the equator of the horizon, which reach the string scale at a coupling $g^2\propto J^5$, \eqref{5dhybrid}. At this point, the black hole transforms into a hybrid (possibly with several spikes). Thin black rings break up into fragments and then become multi-string states. At moderate spins, they can also develop fast elastic instabilities (not shown) and then spin down to move into the Kerr regime.}
        \label{fig:BHStr5}
\end{figure}

We illustrate this in figure~\ref{fig:BHStr5}.
\paragraph{Kerr black holes: $J<S$.} The picture is the same as proposed in $D\geq 6$.

\paragraph{Ultraspinning black holes: $J>S$.} There are two kinds of black holes in this range: thin black rings and MP black holes close to extremality. 

Thin black rings are unstable and will evolve either through elastic instabilities (similar to bar instabilities) at moderate spins, or fragmentation at higher spins. We expect the same picture as in $D\geq 6$: either the formation of a Kerr-regime black hole or the breakup into several of them, in both cases followed by an adiabatic transition into string balls.

Five-dimensional MP black holes close to extremality differ significantly from the near-extremal four-dimensional Kerr solution in that they have $J>S$. They also differ from ultraspinning black holes in $D\geq 6$ in that they are dynamically stable. Their main peculiarity is that they develop large curvatures near their equatorial rim, while the curvature near the poles is much smaller. As a consequence, we expect that the horizon first grows a stringy structure near the equator, and then, through superradiant amplification, the system evolves into a hybrid as in figure~\ref{fig:hybrids}.

The telltale sign is the Kretschmann scalar computed in \eqref{5Keq}. When the black hole is close to extremality, its equatorial value
\begin{align}
    \mathcal{K}\big|_{\rm equator}=\frac1{g^{8/3}\ell_s^4}\frac{J^{20/3}}{S^8} 
\end{align}
signals that a transition to a stringy regime should occur when the coupling reaches the value
\begin{align}\label{5dhybrid}
    g^2=\lp \frac{J}{S}\rp^5 \frac1{S}\,,
\end{align}
(cf.~\eqref{gusbh}). Then, when $g$ is lowered to this value, the black hole evolves into a hybrid. The string-rod component in it emits gravitational radiation and shrinks, but this is slow enough that the system can smoothly reach the correspondence value \eqref{gcorrS} where the black hole component becomes a string ball. Thus the correspondence evolution is smooth and ends on a wiggly string rod.

Five-dimensional MP solutions so close to extremality that their entropy is $S=\ord{1}$ are not amenable to a correspondence of the kind we are studying, but their many intriguing properties make them worth separate attention.\footnote{See \cite{Sheikh-Jabbari:2011sar} for one approach in this direction.}

\bigskip We have managed to provide a picture of the correspondence that encompasses all the black holes that were described in section~\ref{sec:bhs} (save for near-extremal Kerr). No new calculational details of the parametric smoothness of the transition are needed. We have shown that the kind of objects---strings or black holes---that make the adiabatic transition at the correspondence have $J\lesssim S$ and therefore the function $M(S,J)$ only differs by $O(1)$ factors from its value for $J=0$. Therefore, the parametric matching works like in the static case.

\section{Dipole black rings and plasmid strings}
\label{sec:dipoleplasmids}

As we mentioned in section~\ref{sec:stringstates}, there are many possibilities for highly degenerate, distinct states of rotating strings. The ones we discussed are part of an interesting larger class with simple profiles in the rotation plane of the form
\begin{equation}
    \mathbf{x}_L= \sqrt{\alpha' J_L}\, (\sin 2\sigma^+,\cos 2\sigma^+)\,,\qquad 
    \mathbf{x}_R= \sqrt{\alpha' J_R}\, (\sin 2\sigma^-,\cos 2\sigma^-)\,. 
\end{equation}
The parameters $J_{L,R}$ are the left and right moving contributions to the total spin, so that $J=J_L+J_R$.
The oscillations in the transverse plane are now constrained by \eqref{classconstraints} to satisfy
\begin{equation}\label{constr3}
    |\partial_\sigma X^k_{L,R}|^2= \alpha'^2 M^2-4J_{L,R}\,,
\end{equation}
giving rise to an entropy
\begin{equation}\label{entLR}
    S \sim \sqrt{\alpha'\left(\frac{M}{2}\right)^2-J_L}+\sqrt{\alpha'\left(\frac{M}{2}\right)^2-J_R}\,,
\end{equation}
as can easily be confirmed by the explicit construction of corresponding quantum states.
Whenever one of the sectors, say the left-moving one, has the maximum angular momentum for a given mass, then it does not contribute to the entropy (to leading order). We call these states \emph{extremal states}. An extremal state has minimal mass and minimal entropy for a given profile and spin. Non-extremal states with the same profile and spin have larger mass and degeneracy due to wiggling in both the left and right-moving sectors.

In section~\ref{sec:stringstates} we focused on profiles for rods, obtained for $J_L=J_R$. They have the largest entropy for given $M$ and $J$ so we argued that they are likely to match stationary or long-lived black holes. But there are other profiles that, although less entropic, can also match known black holes.

\paragraph{Plasmid strings.} 
States with
\begin{equation}
    J_R=0\,,\qquad 0<J_L=J\leq \frac{\alpha'}4 M^2\,,
\end{equation}
can be described as circular strands of string with a wiggly structure along the circle. They have been dubbed `plasmid strings' in \cite{Blanco-Pillado:2007eit}, who studied the extremal states with $J=\alpha' M^2/4$. Here we generalize them to non-extremal states, with a spin bounded by half the Regge limit
\begin{equation}\label{halfregge}
    J\leq \frac{J_\text{Regge}}2\,.
\end{equation}
The radius of the plasmid circle is
\begin{equation}\label{plasmidR}
    R=\sqrt{\alpha' J}\,,
\end{equation}
and the transverse oscillations $X^k$ yield 
\begin{equation}\label{plasmidS}
    S \sim \frac12\sqrt{\alpha'}M+\sqrt{\frac{\alpha'}4 M^2-J}\,.
\end{equation}

The circular profile makes these states special because, when self-gravity is turned on, the rotation in the $(1,2)$ plane will not radiate gravitational waves. The radiation from these states is thermal, coming from the random wiggly structure. So we expect them to correspond, at large coupling, to stationary black holes.

Before making this connection, let us slightly generalize these states to
\begin{equation}
    \mathbf{x}_L= \sqrt{\frac{\alpha' J}{n_w}}\, (\sin 2n_w\sigma^+,\cos 2n_w\sigma^+)\,, \qquad \mathbf{x}_R=0\,.
\end{equation}
Now the string winds around the circle $n_w\geq 1$ times before closing in on itself. This changes the angular momentum bound to
\begin{equation}\label{reggenw}
    J\leq \alpha'\frac{M^2}{4 n_w}\,,
\end{equation}
the entropy formula to
\begin{equation}\label{plasmidSnw}
    S \sim \frac12\sqrt{\alpha'}M+\sqrt{\frac{\alpha'}4 M^2-n_w J}\,,
\end{equation}
and the radius of the plasmid circle to,
\begin{equation}\label{plasmidRnw}
    R=\sqrt{\frac{\alpha' J}{n_w}}\,.
\end{equation}
As explained in \cite{Blanco-Pillado:2007eit}, this relation follows from the balance between the centrifugal force and the string tension, which sets the radius to the `self-dual' value where
\begin{align}\label{extplasmid}
    \frac{J}{R}=\frac{R n_w}{\alpha'}\,.
\end{align}
When the plasmid is extremal and saturates \eqref{reggenw}, its entropy is
\begin{equation}\label{extplasmidSnw}
    S \sim \sqrt{n_w J}\,.
\end{equation}

When the string coupling grows, the plasmid will maintain its circular profile\footnote{When the ring is thin, the circle will slightly shrink due to its gravitational self-attraction \cite{Blanco-Pillado:2007eit}.} but the wiggly structure will collapse and form a horizon. The result is a \emph{dipole black ring} \cite{Emparan:2004wy}. This is a solution whose only asymptotic conserved charges are $M$ and $J$, but it is nevertheless not neutral: it carries an electric dipole of the three-form field strength $H_{(3)}$ that fundamental strings couple to. This dipole is proportional to the winding number around the circle, $n_w$.  

\paragraph{Dipole black rings.} Solutions for dipole black rings are known in exact form only in $D=5$ \cite{Emparan:2004wy}, but they can easily be constructed approximately in the large-radius limit in any higher $D$ \cite{Blanco-Pillado:2007eit,Caldarelli:2010xz,Emparan:2011hg}. We will work in this regime but, for simplicity, we will discuss explicitly only the case of $D=5$, the other ones differing only by order one numerical factors.

The dipole black ring can be regarded as a closed loop of fundamental string with winding and momentum charges adjusted to satisfy the condition of mechanical equilibrium between tension and centrifugal force. The solutions can be parameterized in terms of the radius of the ring circle $R$, the $S^2$ horizon radius $r_0$, and the `rapidities' (or boost angles) $\eta$ and $\alpha$ that characterize the momentum and winding. It terms of these, the physical magnitudes are
\begin{align}
    M&= R\frac{\pi r_0}{4G}\lp \cosh 2\eta+\cosh 2\alpha +2\rp\,,\\
    J&= R^2 \frac{\pi r_0}{4G}\sinh 2\eta\,,\\
    n_w&=\alpha'\frac{\pi r_0}{4G}\sinh 2\alpha\,,\\
    S&=R\frac{2\pi^2 r_0^2}{G}\cosh\eta\cosh\alpha\,.
\end{align}
The condition of equilibrium for the ring was found in \cite{Emparan:2004wy,Emparan:2011hg} to be
\begin{align}\label{equildip}
    \sinh\eta =\cosh\alpha\,.
\end{align}
Observe that the velocity of the ring is bounded below, $\sinh\eta\geq 1$. The minimum is reached for the neutral black ring, and when the winding charge grows larger, the tension grows and the rotation must accordingly increase. There is a maximum, namely the extremal limit, which is obtained when $\alpha, \eta\to\infty$ and $r_0\to 0$ while $r_0 e^{2\alpha}\simeq r_0 e^{2\eta}$ remain finite. In this limit we have
\begin{align}
    \alpha' M^2=4 n_w J\,,\qquad R=\sqrt{\frac{\alpha' J}{n_w}}\,,
\end{align}
which exactly reproduce the expressions \eqref{reggenw} and \eqref{plasmidRnw} for extremal plasmid string states. The entropy vanishes in this limit and the horizon becomes singular, but this means that stringy effects must be considered to resolve the singularity and obtain a non-zero entropy.

Away from extremality, we can solve for the ring radius to find
\begin{align}
    R^2=\frac{\alpha'J}{n_w}\frac{\sinh 2\alpha}{\sinh 2\eta}
    =\frac{\alpha'J}{n_w}\sqrt{1-\frac1{2\cosh^2\eta}}\,.
\end{align}
This is
\begin{align}
    \frac{R^2}{\ell_s^2}\sim \frac{J}{n_w}\,,
\end{align}
up to factors of order one, for all possible values of $\eta$. We take $J> n_w$ so that the circle radius $R$ is larger than $\ell_s$.
Since the curvature of the ring is dominated by the size of the $S^2$, the correspondence to a string state will occur when $r_0=\ell_s=\sqrt{\alpha'}$. 

Let us now consider the entropy. Passing to string units, we can write it, for all values of the rapidity $\eta$ and up to factors of order one, as 
\begin{align}
    S = \frac{R r_0^2}{g^2\ell_s^3}\cosh 2\eta\,.
\end{align}
Under the same assumptions we also have
\begin{align}\label{plasmSMJ}
    S= \frac{r_0}{\ell_s}\sqrt{\alpha'}M\gtrsim \frac{r_0}{\ell_s} \sqrt{n_w J}\,.
\end{align}
Therefore, when $r_0=\ell_s$ the entropy of the plasmid \eqref{plasmidSnw} is reproduced within the accuracy of the correspondence.\footnote{A similar argument was used in \cite{Horowitz:1996nw} for black holes with fundamental string charge.} The agreement of entropies in the extremal limit \eqref{extplasmidSnw} was verified in \cite{Blanco-Pillado:2007eit}, but we now see it holds more generally. We have shown a matching of entropies, but we could equally well use \eqref{plasmSMJ} to argue that the masses of the plasmids and the dipole rings match when we lower $g$ to reach $r_0=\ell_s$ while keeping $S$, $J$, and $n_w$ fixed. 

The value of $g$ at this correspondence is 
\begin{align}\label{higherg}
    g^2=\frac{Re^{2\eta}}{\ell_s}\frac1{S}=\frac{Je^{2\eta}}{S}\frac1{S}\,,
\end{align}
which generalizes the neutral black ring result \eqref{gring}. This coupling is larger than for Kerr-like black holes with the same entropy since, as we noted, when we lower $g$ keeping the entropy fixed, a thin ring reaches the string scale on its $S^2$ earlier than a round, Kerr-like black hole. The boost factor enhances the effect.

Dipole black rings close to extremality are not expected to suffer from GL or elastic instabilities, so the correspondence transition at \eqref{higherg} is relevant. 
However, far from extremality the GL instabilities will kick in. The winding charge of a dipole ring prevents it from fragmenting, but it will still develop bulges that radiate spin away until, eventually, the ring collapses into a Kerr black hole and the dipole disappears.

\section{Multiple angular momenta}
\label{sec:MultipleJ}

The main elements of the correspondence for single-spin states are easily extended to multiple angular momenta $J_i$, 
\begin{align}
    i=1,\dots, \left\lfloor \frac{D-2}{2}\right\rfloor\,.
\end{align}

String states with multiple spins and large degeneracies are readily constructed with a simple generalization of the solutions in section~\ref{sec:stringstates}: just add an adequate profile of the string in other rotation planes. Our explicit construction, where there is no random wiggling in any of the rotation planes, may eventually consume too many dimensions for rotation and yield significantly lower entropies than generic string states, but we will ignore this issue.

For black holes, even if the possibilities for shapes, topologies, and dynamics grow bewilderingly complex as more angular momenta are added, the useful broad brush picture that we gave in section~\ref{sec:bhs} has a straightforward extension. The Kerr regime consists of black holes where all the angular momenta are moderate,
\begin{align}
   \forall i\quad J_i <S \,.
\end{align}
These black holes are round-shaped and (again glossing over details near the bound) are also stable and unique.
Black holes where at least one spin is large 
\begin{align}
   \exists\, i:\quad  J_i >S 
\end{align}
are ultraspinning. These are elongated along the ultraspinning rotation planes. Besides MP black holes there are several other ultraspinning black hole phases, which include other shapes and horizon topologies. These are expected to be unstable as we discussed earlier, with bar/elastic instabilities and fragmentation instabilities. The exception are black holes that share the property of the single-spin MP solution in $D=5$ of having an extremal limit with a naked curvature singularity of zero area. This happens for the extremal limits of odd-$D$ solutions with $J_1=0$ and $J_{i\neq 1}\neq 0$. The black holes close to these singular extremal limits, with
\begin{align}
    J_1\lesssim S < J_{i\neq 1} \qquad \textrm{in odd $D$,}
\end{align}
are ultraspinning according to our definition\footnote{But not in the conventional definition of `ultraspinning', where we fix the mass instead of the entropy. In that case, if we require all the non-zero $J_i$ to be of a similar order, these spins cannot be arbitrarily large.}, but may be stable, like the 5D single-spin solutions, especially if all the non-zero $J_i$ take similar values. They have localized large curvatures and should evolve to string-black hole hybrids.

With these elements in hand, little imagination is needed to extend the picture of the correspondence in section~\ref{sec:Correspondence}.

The one outlier in our account was the four-dimensional near-extremal Kerr black hole. In $D\geq 5$, black holes with all angular momenta turned on
\begin{align}
   \forall i\quad J_i \neq 0 \,
\end{align}
admit extremal limits with regular, zero-temperature horizons of non-zero area. However, when these black holes are ultraspinning they are expected to be unstable. This includes not only some of the MP extremal black holes, but also all the thin extremal black rings and black holes of other topologies. The evolution of the instability, either through spin radiation or fragmentation, will end up in black holes finitely away from extremality, which then fall into the main evolution picture.

The near-extremal Kerr puzzle reappears only for regular extremal black holes that are stable and without large variation in horizon curvatures, hence not ultraspinning, such that
\begin{align}
   \forall i\quad  J_i \simeq S \,.
\end{align}
These have long, stable throats with a temperature that is too low to match a string ball at the correspondence, even if the entropy of both is the same. They do not fall within our scheme.

We close with an intriguing observation. Five-dimensional extremal neutral black holes with two non-zero angular momenta admit a microscopic description that exactly accounts for their Bekenstein-Hawking entropy \cite{Emparan:2006it}. The picture can actually be extended to the extremal Kerr solution in four dimensions \cite{Horowitz:2007xq}. The microscopic degrees of freedom in these cases are most conveniently identified and counted in a U-dual frame where the black hole has identifiable D-brane charges (corresponding to four stacks of D3-branes wrapped along the diagonals of $T^2\times T^2\times T^2$). The entropy is a U-duality invariant, and thus the counting of states extends to the U-frame where the black hole is a rotating neutral solution. However, this frame is not a string theory but the eleven-dimensional M-theory, where we do not expect a correspondence to string states. Thus, the nature of the microscopic degrees of freedom of these neutral black holes remains particularly enigmatic.

\bigskip We have now completed our picture of the correspondence when rotation is present. We turn to discuss other perspectives on the transition between black holes and strings (with rotation or not). 

\section{Correspondence in evaporation}
\label{sec:evapcorr}

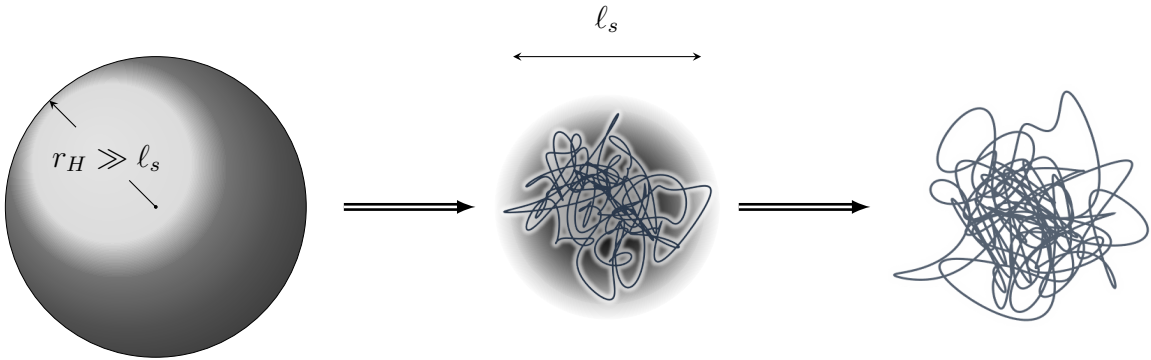
\begin{figure}
	\centering
	\begin{tikzpicture}
    \begin{scope}[shift={(2,2.5)}]
		\begin{scope}
            \shade[ball color = black!55] (0,0) circle (2);
            \draw[black, line width = 0pt] (0,0) circle (2);
            \draw[black] (0,0) -- (-0.5*0.707,0.5*0.707);
            \draw[black, ->, >=stealth] (-1.5*0.707,1.5*0.707) -- (-2*0.707,2*0.707);
            \draw[black, fill=black, anchor=center] (0,0) circle (0.5pt);
        \end{scope}
        \node[black, anchor=center] at (-0.65, 0.65) {\large $r_H \gg \ell_s$};
    \end{scope}
    \draw [line width=1pt, double distance=1pt,
             arrows = {-latex[width=3pt 4 1]}] (4.5,2.5) -- (6.25,2.5);
		\begin{scope}[shift={(8,2.5)}]
			     \begin{scope}
                     \draw[black, <->, >=stealth] (-1.25,2) -- (1.25,2);
                     \node[black, anchor=center] at (0, 2.5) {\large $\ell_s$};
                    \tikzset{myCircle/.style={black!100, path fading=fade out,}} 
                      \fill[myCircle] (0,0) circle (1.5);
                       \tikzset{myCircle/.style={black!100, path fading=fade out,}} 
                      \fill[myCircle] (0,0) circle (1.3);
                    \node[anchor=center] at (0,0) 
                    {\includegraphics[scale =0.3]{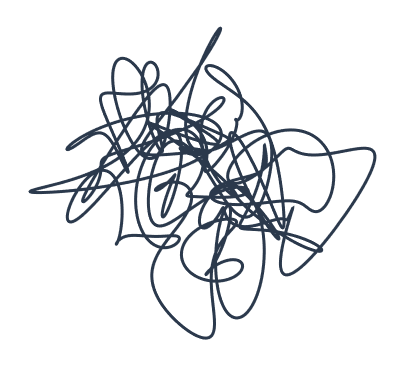}};
            \end{scope}
		\end{scope}
    \draw [line width=1pt, double distance=1pt,
             arrows = {-latex[width=3pt 4 1]}] (9.75,2.5) -- (11.5,2.5);
		\begin{scope}[shift={(13.5,2.5)}]
            \node[anchor=center] at (0,0) 
			  {\includegraphics[scale =0.35]{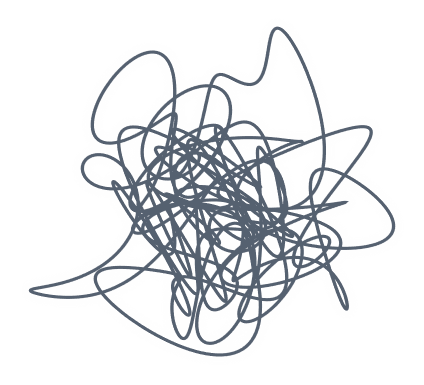}};
		\end{scope}
	\end{tikzpicture}
	\caption{\small An evaporating black hole evolves into a string ball when its radius reaches the string scale $\ell_s$, which corresponds to a mass $M=M_s/g^2$.
	}
	\label{fig:Evap}
\end{figure}

We have put forward the passage of a long dilaton wave as a physical realization of the transition between black holes and strings, but the correspondence also plays an important role in the evaporation of a black hole \cite{Bowick:1985af}. In this case, the coupling $g$ is fixed but the mass decreases as the black hole emits Hawking radiation, and it is natural to expect that when the black hole reaches the string scale, it transitions into a string ball (see figure~\ref{fig:Evap}).

\subsection{Approximate adiabaticity}\label{subsec:evapadiabat}

In contrast to the dilaton wave scheme, in the evaporation scenario we do not get to choose the evolution rate.\footnote{Some control can be gained by adjusting the masses, spins, and number of fields that can be radiated, but this can lead to rather baroque scenarios, possibly in the swampland outside of string theory.} Nevertheless, when $g$ is small it occurs almost adiabatically. The reason is that when the correspondence is reached, \eqref{gcorrM}, the black hole mass
\begin{align}\label{corrM}
    M=\frac{M_s}{g^2}=g^{-2\frac{D-3}{D-2}}M_P
\end{align}
is still much larger than the Planck mass, so its temperature is still very low. We usually think that the evaporation of a black hole goes very fast in its last stages, but if the coupling $g$ is small, then, when the black hole transits into a string ball, the evolution is still very slow in Planck units,
\begin{align}
    \Gamma_\mathrm{corr}=g^{\frac2{D-2}}\ell_P^{-1}\ll \ell_P^{-1}\,.
\end{align}
At this moment, the black hole still has a large entropy, $S\sim 1/g^2\gg 1$, and if the transition occurs during a few string-units of time, the loss of entropy through radiation emission will only be $|\delta S|=\ord{1} \ll S$, so we can regard it as an almost adiabatic evolution.

Therefore, the correspondence transition is physically realized in the evaporation process. However, this way of viewing it has two shortcomings compared to the dilaton wave scheme. First, it does not incorporate a natural way to revert the transition and have the string anti-evaporate into a black hole. The second is that the evolution cannot be controlled by the thought-experimentalist but is governed by the physics of Hawking emission. Especially in the correspondence for evaporating rotating black holes, it is very hard to adjust the change of not only the mass but also the spin. However, the transition remains nearly adiabatic since for a large black hole the spin loss through quantum radiation is also slow in Planck units, even near extremality, as we discussed in section~\ref{subsec:srad}.

\subsection{Page time and the island-string correspondence}

In the picture above we are implicitly assuming that the black hole transits into a highly massive but pure state of the string of the kind studied in section~\ref{sec:stringstates}, at least to a good approximation. This is appropriate if the black hole, at the moment when it reaches the correspondence transition, has only evaporated a small fraction of its initial mass---that is, the black hole is still far from the Page time. If we assume that black hole evaporation is a unitary process, then this black hole will have some entanglement with its early radiation, but its deviation from purity will be small. If it is in this state when its mass reaches the value \eqref{corrM}, the string that it transforms into will also be approximately a pure state.

If, instead, the black hole at the correspondence is past its Page time, then it will be fully entangled with the radiation it has emitted before.  The adiabaticity of the transition implies that the string state after the correspondence will also be fully entangled with the external radiation system. The string will not correspond to a single-string pure state, but rather to a state with a large entanglement with distant low-energy radiation. This may be described as a multi-string state---one very massive string plus many massless strings---but a simpler way of modeling it is to consider a two-string system: we double the Hilbert space of a string and construct a state where each of the oscillators of the string is in a Bell-pair state with an oscillator of the copied string, in a sort of `thermofield-double string'\footnote{This is not exactly the same as a thermo-worldsheet-field double since the oscillators need not carry Boltzmann weights. It is even less a thermo-string-field double of closed string field theory.}. The mathematical construction is simple, but observe that it is also possible to physically realize it: collect all of the radiation that has been emitted and put it in a box small enough that it reaches the Hagedorn point where it makes a transition from a massless radiation gas to a long string. The resulting state of the whole system will be well described (possibly after appropriate distillation) as the massive doubled-string state.

In this manner, we obtain a physical realization of how the island phase in post-Page-time evaporating black holes has a correspondence to `island strings'.

\section{Tracking states and ensembles}
\label{sec:redux}
Since we have been led to introduce new ingredients in the correspondence between black holes and fundamental strings, before we conclude it is appropriate that we reexamine how it works.

The central idea is to follow an ensemble of states as the coupling $g$ is varied. 
Let us consider first how this is done for BPS systems, such as those made of supersymmetric \mbox{D-brane} configurations. There, one starts with the ensemble of BPS states for a given number of branes $N$ at zero coupling. The system is in a stringy or a classical gravitational phase according to whether the 't~Hooft coupling $\lambda=g N$ is small or large.\footnote{Depending on the specific system, $\lambda$ may have different powers of $g$ and $N$, or even several $N_i$.} The index that provides a measure of the degeneracy of BPS ground states is a protected quantity that should not depend on continuous parameters, and in the large $N$ limit is expected to coincide with the exponential of the entropy.  To follow a set of BPS states, we keep $N$ fixed as we vary $g$.\footnote{More generally, the entropy also depends on the spin $J$, which enters in $S(N,J)$ in a similar way as $N$---e.g., shifting the oscillator levels of the CFT---so it must also remain fixed. Recall that $J$ is an adiabatic invariant in classical and quantum mechanics.} The BPS states for given $N$ will mix only among themselves and thus they will remain within the set of states that we started with. That is, the entropy $S(N)$ is fixed, while the mass $M(g,N)$ is renormalized (e.g., for D-branes $M=N M_s/g$, that is, the mass in string units is renormalized at tree level, although not at loop level). For BPS states this renormalization is simple enough that the mass of the string states at $\lambda=0$ can be exactly extrapolated to the mass of the BPS black hole with the same brane numbers $N$ (i.e. charges) at $\lambda\to\infty$. Thus the parameters of the system change smoothly at the correspondence where $\lambda\simeq 1$. The states are stable, so we can vary $g$ as slowly as desired and the process is strictly adiabatic.

Now consider our system, where, as we explained in section~\ref{sec:ReviewStatic}, the parameter 
\begin{align}
    \lambda = g^2 S
\end{align}
plays the role of the 't~Hooft coupling. Following the BPS example, we keep $S$ fixed as $g$ varies. `Fixed entropy' is not any of the usual thermodynamic ensembles, so let us clarify what is the set of states we are following. Again, this is clearest on the string side at $g=0$: we consider configurations of a single free string with different oscillators, but with the same total oscillator number, hence the same $M$, and also the same $J$. This is a microcanonical ensemble of states for a single free string. 

As $g$ grows, interactions renormalize the mass of the string states. Nevertheless, we can expect that, as long as $g$ remains small, the typical states that we start with at $g=0$ will receive essentially the same mass shifts, since these are created by the Newtonian self-gravity of a ball of string, which all these states resemble. Then $S(M,J)$ will remain very approximately the same function, and we can adiabatically track the set of states that we started with. We may think that we are following a sequence of microcanonical ensembles of a single string with given $M(g)$ and $J$. However, we saw that this notion cannot be exact since at any finite~$g$ the string states become unstable and decay in a finite time. We no longer have a strict single-string ensemble, and not even a set of stationary states. The appropriate notion here is the Goldilocks approximately adiabatic evolution introduced in section~\ref{subsec:goldilocks}. We have argued that spinning string states fit within it: they decay slowly enough that they can smoothly evolve from $g=0$ until the correspondence at $g^2 =1/S$, with entropy changes that remain controllably small. 

The situation is less clear in the black hole phase since we do not have a microscopic picture of the states---indeed, the point of the correspondence is to supply that picture, if only at the transition point. It seems natural to still keep the entropy fixed when taking $\lambda>1$ and follow the adiabats with the black hole size and area fixed in Planck units. As we explained in sec.~\ref{sec:ReviewStatic}, the latter is what occurs when $g$ varies by sending a dilaton wave.

It is in the black hole regime that rotation complicates the correspondence picture the most.  As $\lambda$ changes from the classical gravitational limit $\lambda\to\infty$ down to $\lambda =1$, black holes with large spins $J>S$ are not expected to evolve slowly enough to smoothly reach the correspondence. Instead, they decay into black bars, hybrids, and multi-black hole configurations before they can make an almost adiabatic transition to a string state. It is the black hole components of the aftermath that smoothly transit into string balls, since they are black holes with moderate intrinsic spins $J\lesssim S$. 

What is, then, the precise ensemble of states that one follows in the black hole phase? This is unclear, but certainly it is not a sequence of strict microcanonical ensembles of gravitational systems, since these ensembles are not precisely defined even at $\lambda\to\infty$. The gravitational configurations that maximize the entropy for fixed $M$ and $J$ in an asymptotically flat space are never stable stationary regular black hole systems, even in the classical limit. They consist of a \emph{static} central black hole, which carries almost all the entropy, and a larger, distant system that carries the spin at the lowest cost in energy, such as far gravitational waves. We have argued that moderately ultraspinning stationary black holes decay into black bars and hybrids that eventually evolve into configurations close to those. 

Faster ultraspinning black holes instead predominantly fragment and never evolve close to the microcanonically preferred configurations. These black holes must be put in correspondence with multi-string states. This would be the case even if it was possible to fine-tune the ultraspinning black hole so extremely that it would not break apart as we lower $\lambda$ from infinity to one. Our claim is that, if this could be done, the pancaked black hole (or thin black ring) would make a transition into a set of multiple strings put together. Since the system is fast-spinning, it is hard to envisage how it could avoid quickly splitting apart. 

Summing up, although in general the correspondence cannot be made as strict as for BPS systems, we believe that the interpretation as a physical evolution of states that necessarily include dynamical factors is a fruitful view.

\section{Outlook}
\label{sec:Summary}

The self-gravity of rotating string states must still be properly included to fill in the details, but the overall picture of the correspondence that we have given covers a lot of ground.

Several key elements have been necessary to resolve the puzzles posed at the beginning of the article:
\begin{itemize}
    \item Dynamical factors, in particular, black hole instabilities and the emission of radiation.
    \item Non-stationary phases: 
    \begin{itemize}
    \item Black bars.
    \item Black hole-string hybrids.
    \item Multi-string states.
    \end{itemize}
    \item Shapes of strings and black holes, and inhomogeneities of horizon curvatures.
\end{itemize}
None of these ingredients featured in the previous discussions of the correspondence, but rotation has forced their inclusion.

All the puzzles pertained to situations with large spins, $J>S$, either in the string side or in the black hole side, and they have been solved. In particular, the apparent difficulty created by the existence of ultraspinning black holes with $J\gg S$ is solved by realizing that they must be thought of as multi-string states.

Still, at $J=S$ a few but significant cases remain outside our scheme. These are the near-extremal Kerr black holes, which include the four-dimensional solution and the higher-dimensional near-extremal rotating black holes that are not ultraspinning. Identifying the stringy degrees of freedom that describe their throats remains an outstanding problem.

\section*{Acknowledgements}

We thank Iosif Bena, Yiming Chen, Veronika Hubeny, Luca Iliesiu, Raghu Mahajan, David Mateos, Mukund Rangamani, Jorge Santos and Yoav Zigdon for conversations. RE is grateful to Gary Horowitz and Rob Myers for early discussions on this topic. 
The work of N\v{C} is supported by the ERC Grant 787320 - QBH Structure.
RE is supported by MICINN grant PID2019-105614GB-C22, AGAUR grant 2017-SGR 754, and State
Research Agency of MICINN through the ``Unit of Excellence María de Maeztu 2020-2023'' award to the Institute of Cosmos Sciences (CEX2019-000918-M).
AP and MT are supported by the European Research Council (ERC) under the European Union’s
Horizon 2020 research and innovation programme (grant agreement No 852386).

\appendix

\section{The correspondence in Einstein frame}
\label{app:einscorr}

We explained in section~\ref{sec:ReviewStatic} that, as the dilaton wave passes by the black hole, its mass and its area in the Einstein frame do not change, but the stringy corrections in the effective action grow. Here we verify that this yields the same correspondence point as \eqref{gcorrS}.

Very schematically, the gravitational effective action \eqref{Ieiframe} is of the form
\begin{align}\label{Iei2}
    I_\textrm{eff}=\frac1{\ell_s^{D-2}}\int \sqrt{-g_E}\,R_E \lp 1+ \ell_s^2 e^{-\frac{4\phi}{D-2}}R_E+\dots\rp\,,
\end{align}
where $R_E$ generically denotes curvatures in the Einstein frame.
We imagine that the wave is longer than the black hole but is localized in a region of spacetime, so $\phi\to 0$ at infinity; in \eqref{Iei2}, we are implicitly setting $g=1$ asymptotically, so $\ell_s=\ell_P$, but this is not essential.

Before the wave arrives, we have
\begin{align}
    S=\lp\frac{r_H}{\ell_P}\rp^{D-2}=\lp\frac{r_H}{\ell_s}\rp^{D-2}\,,
\end{align}
and the curvature at the horizon is
\begin{align}\label{RES}
    \ell_s^2 R_E =\frac{\ell_s^2}{r_H^2}=S^{-\frac{2}{D-2}}\,.
\end{align}
This is small, and remains constant as the wave passes, but when $\phi$ decreases, the stringy corrections in \eqref{Iei2} are locally enhanced, such that when
\begin{align}
    \ell_s^2 e^{-\frac{4\phi}{D-2}}R_E=1
\end{align}
their effects will not be negligible anymore. Given \eqref{RES}, this happens when
\begin{align}
    e^{2\phi}=\frac1{S}\,,
\end{align}
which coincides with \eqref{gcorrS}.

\section{Higher-dimensional black holes}
\label{app:hidbhs}

Here we collect some results about MP black holes and black rings in several dimensions, keeping all the numerical factors that were omitted in the main text. For more detail, see \cite{Myers:1986un,Emparan:2008eg}. 

\paragraph{Physical magnitudes of MP black holes.}

The coordinate position of the horizon, $r_H$, for the single-spin MP black holes in any $D\geq 4$, is the largest real root of
\begin{align}
    r_H^2+a^2=\frac{\mu}{r_H^{D-5}}\,,
\end{align}
where $\mu$ and $a$ give the mass and spin as
\begin{align}\label{mua}
    \mu=\frac{16\pi}{(D-2)\Omega_{D-2}}GM\,,\qquad a=\frac{D-2}{2}\frac{J}{M}\,.
\end{align}
The entropy is
\begin{align}
    S=\frac{\Omega_{D-2}}{4G}\, r_H^{D-4}(r_H^2+a^2)\,.
\end{align}
We can then write 
\begin{align}\label{rhSM}
    r_H= \frac{D-2}{4\pi}\frac{S}{M}\,,
\end{align}
and
\begin{align}
    \frac{r_H}{a}=\frac{S}{2\pi J}
\end{align}
(valid in all dimensions). Then
\begin{align}\label{GMSJ}
    G M^{D-2} = \frac14 \lp \frac{D-2}{4\pi}\rp^{D-2}\,\Omega_{D-2}\, S^{D-5}\lp S^2 + 4\pi^2 J^2\rp\,,
\end{align}
c.f.~\eqref{MPD}.

The temperature and angular velocity are
\begin{align}
    T_H=\frac1{4\pi}\lp \frac{2r_H^{D-4}}{\mu}+\frac{D-5}{r_H}\rp\,,\qquad 
    \Omega_H=\frac{a}{r_H^2+a^2}\,.
\end{align}
Their explicit expressions in terms of $S$ and $J$ are in general complicated. In section~\ref{subsec:BHSt6D} we will use the result for $D=4$, where the temperature is
\begin{align}\label{TOmSJ}
    T_H=\frac1{4\sqrt{\pi}\,\ell_P}\frac{S^2-4\pi^2 J^2}{S^{3/2}\sqrt{S^2+4\pi^2 J^2}}\,,
\end{align}
with $\ell_P=\sqrt{G}$.

\paragraph{Black rings.} For the black ring in $D=5$, exact expressions are most conveniently given in parametric form,
\begin{align}
    GM^3=\frac{27}{64\pi}S^2\frac1{\nu(1-\nu)}\,,\qquad J^2=\frac1{16\pi^2}S^2\frac{(1+\nu)^3}{\nu^2(1-\nu)}\,,
\end{align}
where the parameter ranges $\nu\in [0,1/2)$ and $\nu\in [1/2,1)$ correspond to thin and fat black rings, respectively. In the thin branch, with $\nu\ll 1$, we find
\begin{align}\label{thinring5}
    GM^3\simeq \frac{27}{16} J\,S\,,
\end{align}
while in the fat branch, when $\nu\to 1$ the extremal value $GM^3=(27\pi/32) J^2$ is approached. 

Thin black rings in $D\geq 6$ at large $J$ satisfy \cite{Emparan:2007wm}
\begin{align}
    GM^{D-2}=\frac1{(4\pi)^{D-4}}\frac{\Omega_{D-3}}8 \frac{(D-2)^{D-2}(D-4)^{\frac{D-4}2}}{(D-3)^\frac{D-3}2}\, J\, S^{D-4}\,,
\end{align}
c.f.~\eqref{BRD}. For $D=5$ we recover \eqref{thinring5}. Detailed analyses in higher dimensions show that this expression remains very approximately valid down to the minimum of $J$ \cite{Armas:2014bia,Dias:2014cia}.

\paragraph{Curvatures.}

The calculation of the Kretschmann curvature 
\begin{align}
    \mathcal{K}=R_{abcd}R^{abcd}
\end{align}
for MP black holes gives very lengthy results but they can be simplified by considering only the polar  and equatorial  values on the horizon. 

For the 5D solution one finds 
\begin{align}
    \mathcal{K}\big|_{\rm axis}&=\frac{24(\mu-4a^2)(3\mu-4a^2)}{\mu^4}\,,\\
    \mathcal{K}\big|_{\rm equator}&=\frac{72\mu^2}{(\mu-a^2)^4}\,.
\end{align}
The sign changes of the polar curvature at values of $a^2<\mu$ are familiar peculiarities of rotating horizons without interest for us (they are narrowly localized in angle, and other curvature invariants may give different results). More relevant is the fact that the equatorial curvature blows up in the extremal limit $a^2\to \mu$. 

Simplifying these results to their parametric dependencies,
\begin{align}
    \mathcal{K}\big|_{\rm axis}&\sim \frac1{\mu^2}\,,\\
    \mathcal{K}\big|_{\rm equator}&\sim \frac{\mu^2}{(\mu-a^2)^4}\,,
\end{align}
and using \eqref{mua}--\eqref{rhSM} to translate into $S$ and $J$, we obtain \eqref{5K0} and \eqref{5Keq}.

For MP black holes in $D\geq 6$ the general expressions are unilluminating but they simplify in the limits of interest. At low spins the curvature is approximately the same as without rotation, that is, \eqref{Kbh} and \eqref{Kstat}. For large spins we find
\begin{align}
    \mathcal{K}\big|_{\rm axis}&\sim \frac1{r_H^4}\,,\\
    \mathcal{K}\big|_{\rm equator}&\sim \frac{\mu^2}{r_H^{2(D-1)}}\,
\end{align}
(near the pole the black hole is well approximated by a black membrane \cite{Emparan:2003sy}). Using \eqref{mua}-\eqref{GMSJ}, these yield \eqref{DK0} and \eqref{DKeq}.

Finally, the curvature of a thin black ring is dominated by the curvature of its small sphere $S^{D-3}$, whose radius is $(S^2/J)^\frac1{D-2}\ell_P$ \cite{Emparan:2007wm}. This immediately gives \eqref{Kring}.

\bibliography{HPbib}

\end{document}